\documentclass[12pt,preprint,number]{elsarticle}
\usepackage{color}
\usepackage{bm}
\usepackage{amsmath}
\usepackage{amssymb}
\usepackage{graphicx}
\usepackage[unicode=true,
 bookmarks=true,bookmarksnumbered=false,bookmarksopen=false,
 breaklinks=false,pdfborder={0 0 1},backref=false,colorlinks=true]
 {hyperref}

\makeatletter

\usepackage{amssymb}
\usepackage{amsmath}
\usepackage{graphicx}
\usepackage{bm}
\usepackage{color}
\usepackage{natbib}

\newcommand{\me}{\mathrm{e}}
\newcommand{\md}{\mathrm{d}}
\newcommand{\mi}{\mathrm{i}}
\newcommand{\mm}{\mathrm{m}}

\newcommand{\mkeV}{\mathrm{keV}}

\newcommand{\fT}{\mathcal{T}}

\newcommand{\fV}{\mathcal{V}}

\journal{CPC}

\makeatother

\begin{document}
\begin{frontmatter}



\title{Gyrokinetic simulation of ITG turbulence with toroidal geometry including
the magnetic axis by using field-aligned coordinates}


\author[ustc]{Zongliang Dai}

\cortext[]{ Corresponding author }

\author[ipp]{Yingfeng Xu}

\author[ipp]{Lei Ye}

\author[ipp]{Xiaotao Xiao}

\author[ustc]{Shaojie Wang\corref{}}

\ead{sjwang@ustc.edu.cn}

\address[ustc]{Department of Engineering and Applied Physics, University of Science
and Technology of China, Hefei, Anhui 230026, China}

\address[ipp]{Institute of Plasma Physics, Chinese Academy of Science, Hefei,
Anhui 230031, China}


\begin{abstract}
Simulation domain in field-aligned coordinates of the electrostatic
gyrokinetic nonlinear turbulence global code, NLT, is extended to
include the magnetic axis. The artificial boundary near the magnetic
axis is replaced by the natural boundary. The singularity at the magnetic
axis in Vlasov solver is treated by considering the spatial relation
of fixed grid points in field-aligned coordinates. A new Poisson's
equation solver is developed, the coefficient matrix of algebraic
equations is derived by using Gauss's theorem. Nonlinear relaxation
test of the ITG turbulence with adiabatic electrons is performed.
The gyrocenter conservation is much improved by including the magnetic
axis in the simulation domain. The zonal field and the radial distribution
of the perturbed electrostatic potential are different from previous
results without the magnetic axis.
\end{abstract}
\end{frontmatter}



\section{Introduction}

\label{Sec:1}

Gyrokinetic simulation is an important tool for investigating properties
of the low frequency turbulence in magnetized plasmas\citep{2010-GARBET-Overview}.
In a tokamak, the low frequency drift wave turbulence and the guiding
center drift motion are inseparable, 3D toroidal geometry is necessary
for gyrokinetic simulation. There are two kinds of space domain selection
in gyrokinetic simulation with 3D toroidal geometry. One is the flux-tube
domain\citep{1995-Beer-FluxTube,1996-Dimits-FluxTube}. This domain
is several correlation lengths wide in both radial and poloidal directions
and extended along the field line. Flux-tube simulations require less
computational cost. But the zonal field in the simulation domain cannot
be evolved self-consistently, and researchers have realized that the
zonal field plays an important role in turbulence nonlinear saturation\citep{1993-Parker-Global,1994-Parker-Comparisons}.
The other one is the global domain\citep{1993-Parker-Global,1995-Sydora-Global,1998-Lin-Zonal}.
It includes most space in a tokamak. The self-consistent evolution
of the zonal field is involved in global simulations. The computational
cost of a global simulation is much more then that of a flux-tube
simulation. With the development of computers, global simulations
have been widely used in research of tokamak plasma physics.

However, the magnetic axis is not included in most global simulations\citep{1998-Lin-Zonal,2003-Candy-GYRO,2006-Grandgirard-GYSELAa,2011-Gorler-GENE}.
Usually, the artificial internal boundary in radial direction is used.
This artificial boundary have an influence on field solver, guiding
center motion and system conservation, which makes some simulation
results difficult to grasp. Recently, in order to prevent particles
from escaping the computational domain, researchers improve the radial
boundaries of GYSELA code\citep{2014-Latu-GYSELAb}, but the magnetic
axis is still not included in simulation and the internal radial boundary
still exist. A new finite element field solver is used in GTC to extend
the simulation domain including the magnetic axis\citep{2018-Hongying-GTC},
but a zero boundary condition at both inner and outer boundaries is
imposed in this new field solver. Difficulty of simulation at the
magnetic axis region mainly comes from the 3D toroidal geometry. Many
equilibrium quantities are functions of the poloidal magnetic flux.
Thus, it is natural to use magnetic coordinates (magnetic flux coordinates
or field-aligned coordinates) in global simulation. Magnetic coordinates
are generalized polar coordinates in the radial-poloidal plane. In
equations of motion, the velocity of poloidal angular coordinate is
singular at the magnetic axis\citep{2007-Jolliet-ORB5c}.

The magnetic axis is included in the simulation domain of the PIC
code ORB5\citep{1999-Parker-ORB5a,1999-Tran-ORB5b,2007-Jolliet-ORB5c}
and the Eulerian code GT5D\citep{2008-Idomura-GT5D}. The finite element
approach is used by these two codes to solve the Poisson's equation\citep{1998-Fivaz-Field,2002-Hatzky-Field,2003-Idomura-GT3D},
a natural boundary condition is imposed at the magnetic axis. In ORB5,
in order to avoid the singularity at the magnetic axis, it is adequate
to use the equivalent cylindrical coordinates, and equilibrium coefficients
needed for the pushing are obtained with linear interpolations\citep{2007-Jolliet-ORB5c}.
And also, the Vlasov solver of the GT5D is treated in cylindrical
coordinates, thus a mapping between cylindrical coordinates and magnetic
flux coordinates is used in the simulation\citep{2008-Idomura-GT5D}.

In this paper, the numerical method to treat the magnetic axis in
the electrostatic gyrokinetic nonlinear turbulence global code, NLT,
by using field-aligned coordinates is presented. NLT is a continuum
code based on the numerical Lie transform method\citep{2016-Ye-NLTa,2017-Xu-NLTb}.
The key idea of the numerical Lie transform method is to decouple
the perturbed motion of the gyrocenter from the unperturbed motion,
and the perturbed distribution function is obtained from the unperturbed
one by using pull-back transform\citep{2012-Wang-I_transform_a,2013-Wang-I_transform_b,2014-Xu-NLT_Orbit,2015-Dai-1D_NLT}.
NLT is mainly composed of four parts: integration along the unperturbed
orbit, pull-back transform, Poisson's equation solver and numerical
filter. Special numerical schemes are adopted in all these parts at
the magnetic axis. The remaining part of this paper is organized as
follows. In Sec.~\ref{Sec:2}, the fundamental equations are introduced,
the previous numerical schemes in NLT and its limitation at the magnetic
axis are reviewed. Sec.~\ref{Sec:3}, computation of the unperturbed
guiding center orbit. Sec.~\ref{Sec:4}, numerical scheme of pull-back
transform at the magnetic axis. Sec.~\ref{Sec:5}, a new Poisson's
equation solver is described. Sec.~\ref{Sec:6}, numerical filter.
Sec.~\ref{Sec:7}, nonlinear relaxation test. Sec.~\ref{Sec:8},
summary and discussion.

\section{Review of the NLT code}

\label{Sec:2}

In this section, the fundamental equations are introduced firstly.
Then the previous numerical schemes in NLT and its limitation at the
magnetic axis are reviewed.

\subsection{Fundamental equations}

\label{Sec:2a}

The gyrocenter distribution function $F\left(\bm{Z}\right)$ satisfies
the gyrokinetic Vlasov equation
\begin{equation}
\frac{\md F}{\md t}\equiv\partial_{t}F+\dot{\bm{X}}\cdot\nabla F+\dot{V}_{\parallel}\partial_{V_{\parallel}}F=0,
\end{equation}
where $\bm{Z}=\left(\bm{X},V_{\parallel},\mu\right)$, and $V_{\parallel}$
is the parallel velocity, $\mu$ is the magnetic moment, $\bm{X}$
is the position of the gyro-center. The gyrokinetic quasi-neutrality
equation in the long-wavelength approximation with adiabatic electron
is\citep{1983-Lee-PD}
\begin{equation}
\nabla\cdot\left(c_{0}\nabla_{\perp}\phi\right)-c_{1}\left(\phi-\left\langle \phi\right\rangle _{FA}\right)=c_{2}\rho_{i,gy}.\label{eq:QN}
\end{equation}
with $c_{0}=\frac{n_{0i}m_{i}}{B^{2}}$, $c_{1}=\frac{e^{2}n_{0e}}{T_{e}}$,
$c_{2}=-e_{i}$. Here $n_{0i}$ and $n_{0e}$ represent the equilibrium
density of ion and electron, respectively, $m_{i}$ is the mass of
ion, $B$ is the equilibrium magnetic field, $T_{e}$ is the temperature
of electron, $e$ and $e_{i}$ respectively represent electric charge
of electron and ion. $\ensuremath{\rho_{i,gy}}$ is the gyrocenter
density of the ion, which is given by 
\begin{equation}
\rho_{i,gy}=\int\md V_{\parallel}\md\mu2\pi B_{\parallel}^{*}\left\langle \delta F\right\rangle _{GA},
\end{equation}
with 
\begin{equation}
B_{\parallel}^{*}=B+\frac{m_{i}V_{\parallel}}{e_{i}}\bm{b}\cdot\nabla\times\bm{b},
\end{equation}
$\bm{b}=\frac{\bm{B}}{B}$. The gyro-average operator $\left\langle \cdot\right\rangle _{GA}$
is defined as
\begin{equation}
\left\langle f\right\rangle _{GA}\left(\bm{r},\mu\right)\equiv\frac{1}{2\pi}\int_{0}^{2\pi}f\left(\bm{X}+\bm{\rho}\left(\mu,\xi\right)-\bm{r}\right)\md\xi,
\end{equation}
$\left\langle \phi\right\rangle _{FA}$ represents the magnetic surface
averaged electrostatic potential. The magnetic surface averaged operator
$\left\langle \cdot\right\rangle _{FA}$ is defined as 
\begin{equation}
\left\langle f\right\rangle _{FA}\equiv\frac{\int_{0}^{2\pi}\md\alpha\int_{0}^{2\pi}\md\theta J_{\bm{X}}f}{\int_{0}^{2\pi}\md\alpha\int_{0}^{2\pi}\md\theta J_{\bm{X}}},\label{eq:FA}
\end{equation}
with $J_{\bm{X}}$ being the space Jacobian.

The equilibrium magnetic field can be expressed in terms of magnetic
flux coordinates $\bm{X}_{f}\equiv(\psi^{*},\theta^{*},\zeta^{*})$
as 
\begin{equation}
\bm{B}=g\left(\psi^{*}\right)\nabla\zeta^{*}+I\left(\psi^{*}\right)\nabla\theta^{*}+g\left(\psi^{*}\right)\delta\left(\psi^{*},\theta^{*}\right)\nabla\psi^{*},
\end{equation}
 where $\psi^{*}$ is the poloidal magnetic flux, $\theta^{*}$ is
the poloidal angle, $\zeta^{*}$ is the toroidal angle. $I\left(\psi^{*}\right)$
and $g\left(\psi^{*}\right)$ represent the toroidal and poloidal
components of the magnetic field in the covariant form, respectively.
For micro-turbulence in tokamak plasmas, the perpendicular wavenumber
is much larger then the parallel wavenumber, $k_{\perp}\gg k_{\parallel}$.
Therefore, field-aligned coordinates $\bm{X}_{l}\equiv(\psi,\theta,\alpha)$
can be used to improve the computational efficiency, where
\begin{align}
 & \psi=\psi^{*},\label{eq:flux2field1}\\
 & \theta=\theta^{*},\label{eq:flux2field2}\\
 & \alpha=q\left(\psi^{*}\right)\theta^{*}-\zeta^{*},\label{eq:flux2field3}
\end{align}
$q\left(\psi^{*}\right)$ is the safety factor. The space Jacobian
is 
\begin{equation}
J_{\bm{X}}\equiv J_{\psi,\theta,\alpha}=J_{\psi^{*},\theta^{*},\zeta^{*}}=\frac{gq+I}{B^{2}}.
\end{equation}
For convenience, $\bm{Z}\equiv\left(\bm{X}_{l},V_{\parallel},\mu\right)$
in the rest part of this paper. 

\subsection{Numerical algorithm}

\label{Sec:2b}

Unlike traditional continuum methods, the process of solving the Vlasov
equation in numerical Lie transform is divided into $2$ sub-processes,
the unperturbed solver and the perturbed solver\citep{2012-Wang-I_transform_a,2013-Wang-I_transform_b}.
The unperturbed solver is treated by integrating along the unperturbed
orbit. The perturbed solver is treated by pull-back transform, which
is equivalent to compute the perturbed orbit. Thus, NLT is mainly
composed of $4$ parts: integration along the unperturbed orbit, pull-back
transform, Poisson's equation solver and numerical filter.

In the first part, $\delta\bar{F}$ and $S_{1}$ are computed\citep{2016-Ye-NLTa},
the former represents the evolution of the perturbed distribution
function $\delta F$ under the equilibrium field, the latter represents
the gauge function of the I-transform\citep{2012-Wang-I_transform_a,2013-Wang-I_transform_b},
\begin{align}
 & \frac{\md_{0}}{\md t}\delta\bar{F}=0,\\
 & \frac{\md_{0}}{\md t}S_{1}=e_{i}\left\langle \phi\right\rangle _{GA},
\end{align}
 the total time derivative $\frac{\md_{0}}{\md t}$ is taken along
the unperturbed orbit. At the beginning of each time step, $S_{1}=0$,
$\delta\bar{F}=\delta F$. Numerically, we can obtain the solution
by using the semi-Lagrangian method and the high dimensional fixed
point interpolation algorithm\citep{2017-Xiao-Interpolation}, 
\begin{align}
 & \delta\bar{F}\left(\bm{Z},t+\Delta t\right)=\delta\bar{F}\left(\bm{Z}+\Delta\bm{Z}_{0}\left(-\Delta t\right),t\right)=\delta F\left(\bm{Z}+\Delta\bm{Z}_{0}\left(-\Delta t\right),t\right),\\
 & S_{1}\left(\bm{Z},t+\Delta t\right)=e_{i}\int_{t}^{t+\Delta t}\left\langle \phi\right\rangle _{GA}\left(\bm{Z}+\Delta\bm{Z}_{0}\left(t-\tau\right),\tau\right)\md\tau,
\end{align}
where 
\begin{equation}
\Delta\bm{Z}_{0}\left(t\right)\equiv\int_{0}^{t}\left\{ \bm{Z},H_{0}\right\} \md t,
\end{equation}
$H_{0}$ is the unperturbed guiding center Hamiltonian, 
\begin{equation}
H_{0}=\frac{1}{2}mV_{\parallel}^{2}+\mu B,
\end{equation}
 $\{\text{},\text{}\}$ is the Poisson bracket, 
\begin{equation}
\left\{ f,g\right\} =\frac{\partial f}{\partial Z^{a}}J^{Z^{a}Z^{b}}\frac{\partial g}{\partial Z^{b}},
\end{equation}
$J^{Z^{a}Z^{b}}$ is the component of the unperturbed Poisson matrix,
which can be expressed as\citep{2017-Xu-NLTb} 
\begin{align}
 & J^{\psi\theta}=-J^{\theta\psi}=J^{\psi^{*}\theta^{*}},\\
 & J^{\psi\alpha}=-J^{\alpha\psi}=qJ^{\psi^{*}\theta^{*}}-J^{\psi^{*}\zeta^{*}},\\
 & J^{\psi V_{\parallel}}=-J^{V_{\parallel}\psi}=J^{\psi^{*}V_{\parallel}},\\
 & J^{\theta\alpha}=-J^{\alpha\theta}=-J^{\theta^{*}\zeta^{*}}-q'\theta J^{\psi^{*}\theta^{*}},\\
 & J^{\theta V_{\parallel}}=-J^{V_{\parallel}\theta}=J^{\theta^{*}V_{\parallel}},\\
 & J^{\alpha V_{\parallel}}=-J^{V_{\parallel}\alpha}=qJ^{\theta^{*}V_{\parallel}}+q'\theta J^{\psi^{*}V_{\parallel}}-J^{\zeta^{*}V_{\parallel}},
\end{align}
with $J^{Z_{f}^{a}Z_{f}^{b}}$ the form of the unperturbed Poisson
matrix component in $\bm{Z}_{f}\equiv\left(\bm{X}_{f},V_{\parallel},\mu\right)$
\begin{align}
 & J^{\psi^{*}\theta^{*}}=-\frac{g}{e_{i}D},\\
 & J^{\psi^{*}\zeta^{*}}=\frac{I}{e_{i}D},\\
 & J^{\psi^{*}V_{\parallel}}=\frac{V_{\parallel}B}{e_{i}D}\partial_{\theta^{*}}\left(\frac{g}{B}\right),\\
 & J^{\theta^{*}\zeta^{*}}=\frac{g\delta}{e_{i}D},\\
 & J^{\theta^{*}V_{\parallel}}=-J^{V_{\parallel}\theta}=\frac{B}{m_{i}D}\left[1-\frac{m_{i}V_{\parallel}}{e_{i}}\partial_{\psi^{*}}\left(\frac{g}{B}\right)\right],\\
 & J^{\zeta^{*}V_{\parallel}}=\frac{B}{m_{i}D}\left\{ q+\frac{m_{i}V_{\parallel}}{e_{i}}\left[\partial_{\psi^{*}}\left(\frac{I}{B}\right)+\partial_{\theta^{*}}\left(\frac{g\delta}{B}\right)\right]\right\} ,\\
 & q'=\frac{\md q}{\md\psi^{*}},\\
 & D=qg+I+\rho_{\parallel}\left(I'g-g'I\right)-\rho_{\parallel}g^{2}\partial_{\theta^{*}}\delta,\\
 & \rho_{\parallel}=\frac{m_{i}V_{\parallel}}{e_{i}B}.
\end{align}

Usually, in the region away from the magnetic axis, $J^{\theta V_{\parallel}}\approx\frac{B}{m_{i}D}$.
However, in the region near the magnetic axis, when $\psi\to0$, $J^{\theta V_{\parallel}}\approx\frac{BV_{\parallel}}{De_{i}}\partial_{\psi}\left(\frac{g}{B}\right)\propto r^{-1}$,
where $r$ is the minor radius. $J^{\theta V_{\parallel}}$ contributes
the velocity of guilding center in $V_{\parallel}$ and $\theta$
directions. The contribution in $V_{\parallel}$ direction is $-J^{\theta V_{\parallel}}\partial_{\theta}H_{0}$,
near the magnetic axis, $\partial_{\theta}H_{0}\propto r$, thus this
contribution is not divergent, but it is difficult to be treated numerically.
The contribution in $\theta$ direction is $J^{\theta V_{\parallel}}\partial_{V_{\parallel}}H_{0}=V_{\parallel}J^{\theta V_{\parallel}}$,
it is divergent; as is mentioned in the introduction, this singularity
is due to the properties of the generalized polar coordinates.

In the second part, $\delta F$ is computed by using pull-back transform\citep{2017-Xu-NLTb}
\begin{align}
 & \delta F=\delta\bar{F}+\delta F_{A}+\delta F_{B},\label{eq:Pull-back_Ff}\\
 & \delta F_{A}=G_{1}^{Z^{a}}\frac{\partial}{\partial Z^{a}}\left(F_{0}+\delta\bar{F}\right),\\
 & \delta F_{B}=\frac{1}{2}G_{1}^{Z^{a}}\frac{\partial}{\partial Z^{a}}\delta F_{A},
\end{align}
 where $\bm{G}_{1}$ is the 1st order generating vector field,
\begin{equation}
\bm{G}_{1}=\left\{ S_{1},\bm{Z}\right\} .
\end{equation}

The main task of this part is to compute numerical differentiations.
Note that the Poisson matrix is needed for computing $\bm{G}_{1}$,
the singularity will also appear at the magnetic axis. Considering
that the pull-back transform is equivalent to the computation of the
perturbed orbit, the problem and the solution of this singularity
are same as that in the first part. 

The gyrokinetic quasi-neutrality equation is solved in the third part\citep{2017-Xu-NLTb}.
$\left\langle \phi\right\rangle _{FA}$ is solved approximately from
the equation by taking the magnetic surface average on both sides
of the Eq.~(\ref{eq:QN}). The partial differential operator in $\alpha$
direction is converted to algebraic operator by the toroidal Fourier
transform. Further, by considering that the $\theta$ direction is
parallel to the magnetic field line in field-aligned coordinates and
$k_{\perp}\gg k_{\parallel}$, we regard the term containing $\partial_{\theta}$
as a correction in numerical, which can be computed iteratively. Thus,
the three-dimensional second-order partial differential operators
reduced to one-dimensional second-order differential operators. In
numerical, $\phi$ is solved iteratively by using the finite difference
method with the Dirichlet boundary conditions.

There are three shortcomings in this field solver. First, $\left\langle \phi\right\rangle _{FA}$
is an approximate solution; Second, the Dirichlet boundary condition
at the inner boundary is not self-consistent; Third, at the magnetic
axis, $\partial_{\theta}=-q\partial_{\alpha}$, thus the term containing
$\partial_{\theta}$ cannot be treated as a correction in numerical.

The last part is the numerical filter. The Fourier filter, zeroing
the coefficients of Fourier components with wavelengths of these components
being less than three times the width of the grid, is adopted in NLT\citep{2017-Xu-NLTb}.
The phase-space density function $J_{\bm{Z}}\delta F$ is filtered
in each time step, where $J_{\bm{Z}}$ is the phase-space Jacobian.
Note that $\delta F$ is very small near the computational velocity
boundary and the damping buffer regions are used near the radial boundaries,
the Gibbs phenomenon has a less impact on the simulation. Thus, although
the boundary conditions in both $\psi$ and $V_{\parallel}$ directions
are not periodic, the Fourier filter is still used in these directions.
Considering that $\theta$ is also non-periodic in the field-aligned
coordinates, we transform $J_{\bm{Z}}\delta F$ from field-aligned
coordinates into magnetic flux coordinates$J_{\bm{Z}_{f}}\delta F_{f}$
to truncate the shortwave in the parallel direction by using the Fourier
filter, with $J_{\bm{Z}_{f}}$ the phase-space Jacobian of $\bm{Z}_{f}$.
$\left(J_{\bm{Z}_{f}}\delta F_{f}\right)_{mn}$ is the Fourier component
of $J_{\bm{Z}_{f}}\delta F_{f}$, $n$ and $m$ are the mode number
in $\zeta^{*}$ and $\theta^{*}$ directions, respectively. Filter
conditions are determined by fixed grid points in field-aligned coordinates,
$\left(J_{\bm{Z}_{f}}\delta F_{f}\right)_{mn}=0$ when $n>N_{\alpha}/3$,
$m<nq-N_{\theta}/3$ or $m\text{>}nq+N_{\theta}/3$, where $N_{\theta}$
and $N_{\alpha}$ are the total number of grid points in the $\theta$
and $\alpha$ directions.

If the magnetic axis is included in the simulation domain, the inner
damping buffer region should not be used any more, the filter in the
radial direction need to be redesigned. And note that the fluctuation
with high wave number in $\theta^{*}$ direction, $k_{\theta}=m/r$,
will not be truncated near the magnetic axis by taking only the Fourier
filter condition

\section{Computation of the unperturbed guiding center orbit near the magnetic
axis}

\label{Sec:3}

The numerical singularity will appear if the orbit near the magnetic
axis is computed by using Hamilton's equations in magnetic coordinates.
For treating this problem, we compute the unperturbed orbit by using
Hamiltonian equations in cylindrical coordinates $\bm{X}_{c}\equiv\left(R,Z,\zeta\right)$
if the radial position of the guiding center is close to the magnetic
axis. This method has been used in previous work\citep{2014-Ye-AxisOrbit}.

$\fT_{l\to c}$ and $\fT_{c\to l}$ represent the coordinate transformation
and inverse transformation between $\bm{Z}$ and $\bm{Z}_{c}\equiv\left(\bm{X}_{c},V_{\parallel},\mu\right)$,
respectively, 
\begin{align}
 & \bm{Z}_{c}=\fT_{l\to c}\left(\bm{Z}\right),\label{eq:Zc2Zl1}\\
 & \bm{Z}=\fT_{c\to l}\left(\bm{Z}_{c}\right).\label{eq:Zc2Zl2}
\end{align}
Hamiltonian equations in cylindrical coordinates are 
\begin{equation}
\frac{\md\bm{Z}_{c}}{\md t}=\left\{ \bm{Z}_{c},H_{c,0}\right\} _{c}\biggl|_{\bm{Z}_{c}=\fT_{l\to c}\left(\bm{Z}\right)},
\end{equation}
where $H_{c,0}$ is the unperturbed Hamiltonian in cylindrical coordinates
\begin{equation}
H_{c,0}\left(\bm{Z}_{c}\right)=H_{0}\left(\bm{Z}\right)=H_{0}\left(\fT_{c\to l}\left(\bm{Z}_{c}\right)\right),
\end{equation}
$\left\{ ,\right\} _{c}$ is the Poisson bracket in cylindrical coordinates,
with components of the Poisson matrix 
\begin{align}
J^{RV_{\parallel}}= & -J^{V_{\parallel}R}=\frac{\bm{B}_{0}^{*}\cdot\nabla R}{m_{s}B_{0\parallel}^{*}}\nonumber \\
= & \frac{1}{m_{s}B_{0\parallel}^{*}}\left(-\frac{1}{R}\partial_{Z}\psi+\frac{m_{s}V_{\parallel}}{e_{s}BR}\left(\partial_{Z}g-\frac{g}{B}\partial_{Z}B\right)\right),\\
J^{ZV_{\parallel}}= & -J^{V_{\parallel}Z}=\frac{\bm{B}_{0}^{*}\cdot\nabla Z}{m_{s}B_{0\parallel}^{*}}\nonumber \\
= & \frac{1}{m_{s}B_{0\parallel}^{*}}\left(\frac{1}{R}\partial_{R}\psi+\frac{m_{s}V_{\parallel}}{e_{s}BR}\left(-\partial_{Z}g+\frac{g}{B}\partial_{Z}B\right)\right),\\
J^{\zeta V_{\parallel}}= & -J^{V_{\parallel}\zeta}=\frac{\bm{B}_{0}^{*}\cdot\nabla\zeta}{m_{s}B_{0\parallel}^{*}}=\frac{1}{m_{s}B_{0\parallel}^{*}}\Bigg[\frac{g}{R^{2}}+\frac{m_{s}V_{\parallel}}{e_{s}BR}\nonumber \\
 & \left(-\frac{1}{R}\partial_{R}\psi-\frac{1}{B}\partial_{R}B\partial_{R}\psi+\partial_{RR}^{2}\psi+\partial_{ZZ}^{2}\psi-\frac{1}{B}\partial_{Z}B\partial_{Z}\psi\right)\Bigg],\\
J^{RZ}= & -J^{ZR}=-\frac{\bm{b}_{0}\cdot\nabla R\times\nabla Z}{e_{s}B_{0\parallel}^{*}}=-\frac{g}{e_{s}B_{0\parallel}^{*}BR},\\
J^{R\zeta}= & -J^{\zeta R}=-\frac{\bm{b}_{0}\cdot\nabla R\times\nabla\zeta}{e_{s}B_{0\parallel}^{*}}=\frac{\partial_{R}\psi}{e_{s}B_{0\parallel}^{*}BR^{2}},\\
J^{Z\zeta}= & -J^{\zeta Z}=-\frac{\bm{b}_{0}\cdot\nabla Z\times\nabla\zeta}{e_{s}B_{0\parallel}^{*}}=\frac{\partial_{Z}\psi}{e_{s}B_{0\parallel}^{*}BR^{2}},
\end{align}
and 
\begin{align}
 & J^{\psi V_{\parallel}}=\frac{V_{\parallel}g}{e_{s}B_{0\parallel}^{*}BR}\left(\partial_{R}B\partial_{Z}\psi-\partial_{Z}B\partial_{R}\psi\right),\\
 & B_{0\parallel}^{*}=B_{0}+\frac{m_{s}V_{\parallel}}{e_{s}B^{2}R^{2}}\left(g\Delta^{*}\psi-g'|\nabla\psi|^{2}\right).
\end{align}
Informations of the unperturbed orbit will be recorded in field-aligned
coordinates by using Eqs.~(\ref{eq:Zc2Zl1}) and (\ref{eq:Zc2Zl2}).

It is worth to point out that the unperturbed orbit is independent
of perturbations, which can be computed only once by using the high-precisional
numerical algorithm in the initial of NLT simulation\citep{2016-Ye-NLTa}. 

\section{Pull-back transform at the magnetic axis}

\label{Sec:4}

$\left(\psi_{i},\theta_{j},\alpha_{k}\right)$ represents the spatial
grid point in NLT, with $i=0,1,\cdots,N_{\psi}-1$, $\psi_{0}=0$,
$\psi_{N_{\psi}-1}=\psi_{b}$, $j=0,1,\cdots,N_{\theta}-1$, $\theta_{0}=-\pi$,
$\theta_{N_{\theta}-1}=\pi-\Delta_{\theta}$, $k=0,1,\cdots,N_{\alpha}-1$,
$\alpha_{0}=0$, $\alpha_{N_{\alpha}-1}=2\pi-\Delta_{\alpha}$. The
grid width in $\psi$, $\theta$, $\alpha$ directions are $\Delta_{\psi}=\psi_{b}/\left(N_{\psi}-1\right)$,
$\Delta_{\theta}=2\pi/N_{\theta}$, $\Delta_{\alpha}=2\pi/N_{\alpha}$
respectively. The distribution function satisfies the scalar invariance
\begin{equation}
\delta F\left(\bm{Z}\right)=\delta F_{c}\left(\fT_{l\to c}\left(\bm{Z}\right)\right),
\end{equation}
thus the pull-back transform at the magnetic axis is computed by using
formulations in cylindrical coordinates to avoid the singularity,
\begin{align}
 & \delta F_{c}=\delta\bar{F}_{c}+\delta F_{c,A}+\delta F_{c,B},\label{eq:Pull-Back_Fc}\\
 & \delta F_{c,A}=G_{c,1}^{Z_{c}^{a}}\frac{\partial}{\partial Z_{c}^{a}}\left(F_{c,0}+\delta\bar{F}_{c}\right),\\
 & \delta F_{c,B}=\frac{1}{2}G_{c,1}^{Z_{c}^{a}}\frac{\partial}{\partial Z_{c}^{a}}\delta F_{c,A},
\end{align}
with
\begin{align}
 & \delta\bar{F}_{c}\left(\bm{Z}_{c}\right)=\delta\bar{F}\left(\bm{Z}\right)=\delta\bar{F}\left(\fT_{c\to l}\left(\bm{Z}_{c}\right)\right),\label{eq:S_Fc}\\
 & S_{c,1}\left(\bm{Z}_{c}\right)=S_{1}\left(\bm{Z}\right)=S_{1}\left(\fT_{c\to l}\left(\bm{Z}_{c}\right)\right),\\
 & \bm{G}_{c,1}=\left\{ \bm{Z}_{c},S_{c,1}\right\} _{c}.
\end{align}
Numerically, the coordinate transform from field-aligned coordinates
to the cylindrical coordinates is usually needed to compute $\partial_{R}$,
$\partial_{Z}$ and $\partial_{\zeta}$ in field-aligned coordinates,
which leads to numerical errors. However, it is not used in NLT. By
noticing the spatial relation of fixed grid points in field-aligned
coordinates, we compute partial derivatives at the magnetic axis in
the $R$ direction of cylindrical coordinates by using values at grid
points $\left(\psi=\Delta_{\psi},\theta=-\pi\right)$ and $\left(\psi=\Delta_{\psi},\theta=0\right)$
in field-aligned coordinates. Similarly, the partial derivatives in
the $Z$ direction of cylindrical coordinates can be computed by using
values of grid points $\left(\psi=\Delta_{\psi},\theta=-\pi/2\right)$
and $\left(\psi=\Delta_{\psi},\theta=\pi/2\right)$ in field-aligned
coordinates. If $N_{\theta}$ is an integer multiple of $4$, the
above $4$ points are all existed grid points on $\psi-\theta$ plane,
$\left(\psi=\Delta_{\psi},\theta=-\pi\right)=\left(\psi_{1},\theta_{0}\right)$,
$\left(\psi=\Delta_{\psi},\theta=-\pi/2\right)=\left(\psi_{1},\theta_{\frac{1}{4}N_{\theta}}\right)$,
$\left(\psi=\Delta_{\psi},\theta=0\right)=\left(\psi_{1},\theta_{\frac{1}{2}N_{\theta}}\right)$,
$\left(\psi=\Delta_{\psi},\theta=\pi/2\right)=\left(\psi_{1},\theta_{\frac{3}{4}N_{\theta}}\right)$.
In nonlinear ITG simulations, $N_{\theta}=16$. Thus, only the 1D
transform of toroidal coordinate from $\alpha$ in field-aligned coordinates
to $\zeta$ in cylindrical coordinates is needed, and this transform
can be computed by using the high-precisional 1D Fourier transform.

Coordinates of a space point $p$ can be expressed as 
\begin{align}
 & \bm{X}_{l}|_{p}=\left(\psi|_{p},\theta|_{p},\alpha|_{p}\right),\\
 & \bm{X}_{f}|_{p}=\left(\psi^{*}|_{p},\theta^{*}|_{p},\zeta^{*}|_{p}\right),\\
 & \bm{X}_{c}|_{p}=\left(R|_{p},Z|_{p},\zeta|_{p}\right),
\end{align}
where
\begin{align}
 & \alpha|_{p}=q\left(\psi^{*}|_{p}\right)\theta^{*}|_{p}-\zeta^{*}|_{p},\label{eq:zeta2alpha}\\
 & \zeta^{*}|_{p}=q\left(\psi|_{p}\right)\theta|_{p}-\alpha|_{p},\label{eq:alpha2zeta}\\
 & \zeta|_{p}=\zeta^{*}|_{p}.
\end{align}
\begin{figure}
\centering{}\includegraphics[width=0.45\textwidth]{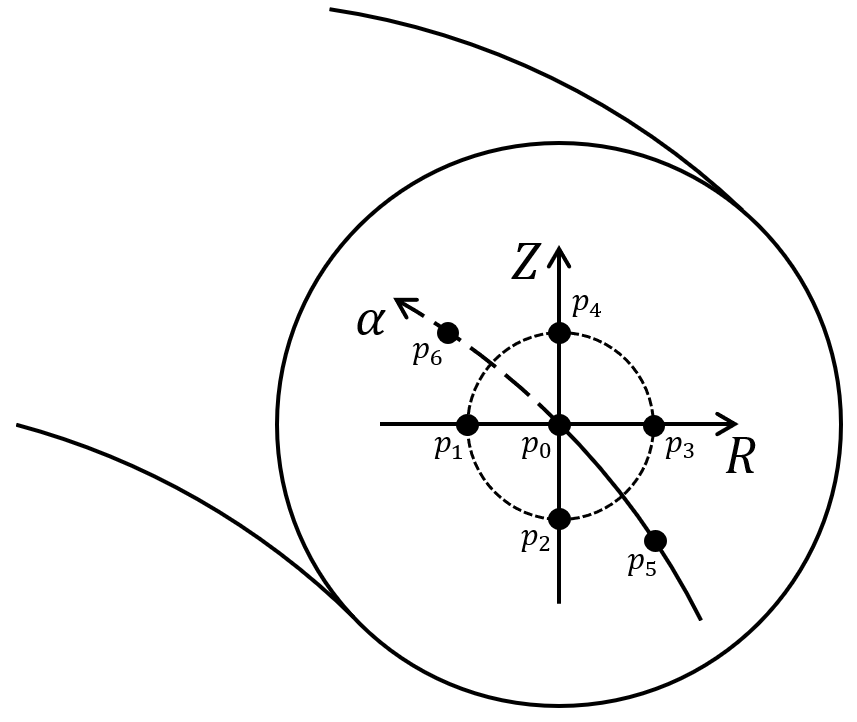}\caption{\label{fig:Pull-back}Partial derivative computation at the magnetic
axis in field-aligned coordinates.}
\end{figure}
As is shown in Fig.~(\ref{fig:Pull-back}), $p_{0}$ is a fixed grid
point at the magnetic axis in field-aligned coordinates. 
\begin{align}
\psi|_{p_{0}} & =\psi_{0},\\
\theta|_{p_{0}} & =\theta_{j},\\
\alpha|_{p_{0}} & =\alpha_{k}.
\end{align}
Thus, in cylindrical coordinates, partial derivatives of any scalar
function $f\left(p\right)$ at $p=p_{0}$ can be numerically computed
by 
\begin{align}
 & \partial_{R}f\left(p_{0}\right)=\frac{1}{R|_{p_{3}}-R|_{p_{1}}}\left(f\left(p_{3}\right)-f\left(p_{1}\right)\right),\label{eq:diravation1}\\
 & \partial_{Z}f\left(p_{0}\right)=\frac{1}{Z|_{p_{4}}-Z|_{p_{2}}}\left(f\left(p_{4}\right)-f\left(p_{2}\right)\right),\label{eq:diravation2}\\
 & \partial_{\zeta}f\left(p_{0}\right)=\frac{1}{\zeta|_{p_{6}}-\zeta|_{p_{5}}}\left(f\left(p_{6}\right)-f\left(p_{5}\right)\right),\label{eq:diravation3}
\end{align}
For avoiding the high dimensional interpolation on $\psi-\theta$
plane, we take 
\begin{align}
 & \bm{X}_{f}|_{p_{1}}=\left(\psi_{1},-\pi,\zeta^{*}|_{p_{0}}\right)=\left(\psi_{1},\theta_{0},\zeta^{*}|_{p_{0}}\right),\\
 & \bm{X}_{f}|_{p_{2}}=\left(\psi_{1},-\frac{\pi}{2},\zeta^{*}|_{p_{0}}\right)=\left(\psi_{1},\theta_{\frac{1}{4}N_{\theta}},\zeta^{*}|_{p_{0}}\right),\\
 & \bm{X}_{f}|_{p_{3}}=\left(\psi_{1},0,\zeta^{*}|_{p_{0}}\right)=\left(\psi_{1},\theta_{\frac{1}{2}N_{\theta}},\zeta^{*}|_{p_{0}}\right),\\
 & \bm{X}_{f}|_{p_{4}}=\left(\psi_{1},\frac{\pi}{2},\zeta^{*}|_{p_{0}}\right)=\left(\psi_{1},\theta_{\frac{3}{4}N_{\theta}},\zeta^{*}|_{p_{0}}\right),\\
 & \bm{X}_{f}|_{p_{5}}=\left(\psi_{0},\theta|_{p_{5}},\zeta^{*}|_{p_{0}}-\Delta\zeta\right),\\
 & \bm{X}_{f}|_{p_{6}}=\left(\psi_{0},\theta|_{p_{6}},\zeta^{*}|_{p_{0}}+\Delta\zeta\right).
\end{align}
By using Eq.~(\ref{eq:alpha2zeta}), we have 
\begin{equation}
\zeta^{*}|_{p_{0}}=q_{0}\theta_{j}-\alpha_{k},
\end{equation}
$q_{i}\equiv q\left(\psi_{i}\right)$. For convenience of computation,
we choose 
\begin{align}
\theta|_{p_{5}} & =\theta|_{p_{6}}=\theta_{j},\\
\Delta\zeta & =\Delta\alpha,
\end{align}
then we obtain 
\begin{align}
 & \bm{X}_{l}|_{p_{1}}=\left(\psi_{1},\theta_{0},\alpha_{k}+q_{0}\left(\theta_{0}-\theta_{j}\right)\right),\\
 & \bm{X}_{l}|_{p_{2}}=\left(\psi_{1},\theta_{\frac{1}{4}N_{\theta}},\alpha_{k}+q_{0}\left(\theta_{\frac{1}{4}N_{\theta}}-\theta_{j}\right)\right),\\
 & \bm{X}_{l}|_{p_{3}}=\left(\psi_{1},\theta_{\frac{1}{2}N_{\theta}},\alpha_{k}+q_{0}\left(\theta_{\frac{1}{2}N_{\theta}}-\theta_{j}\right)\right),\\
 & \bm{X}_{l}|_{p_{4}}=\left(\psi_{1},\theta_{\frac{3}{4}N_{\theta}},\alpha_{k}+q_{0}\left(\theta_{\frac{3}{4}N_{\theta}}-\theta_{j}\right)\right),\\
 & \bm{X}_{l}|_{p_{5}}=\left(\psi_{0},\theta_{j},\alpha_{k}+\Delta_{\alpha}\right)=\begin{cases}
\left(\psi_{0},\theta_{j},\alpha_{k+1}\right), & k\ne N_{\alpha}-1\\
\left(\psi_{0},\theta_{j},\alpha_{1}\right), & k=N_{\alpha}-1
\end{cases}\\
 & \bm{X}_{l}|_{p_{6}}=\left(\psi_{0},\theta_{j},\alpha_{k}-\Delta_{\alpha}\right)=\begin{cases}
\left(\psi_{0},\theta_{j},\alpha_{k-1}\right), & k\ne1\\
\left(\psi_{0},\theta_{j},\alpha_{N_{\alpha}-1}\right). & k=1
\end{cases}
\end{align}

By using the spatial relation of fixed grid points in field-aligned
coordinates and the high-precisional 1D Fourier transform, partial
derivatives in $R$, $Z$ and $\zeta$ directions of cylindrical coordinates
are computed at the magnetic axis in field-aligned coordinates. Further,
pull-back transform for computing the perturbed distribution function
at the magnetic axis is computed by using Eqs.~(\ref{eq:Pull-Back_Fc})
and (\ref{eq:S_Fc}).

\section{Poisson's equation solver}

\label{Sec:5}

In Ref.~\citep{2004-Birdsell-Book}, Gauss's law is used to discretized
2D Poisson's equation in polar coordinates, thus the artificial boundary
condition at the polar is not needed. We extend this algorithm from
2D polar coordinates to 3D field-aligned coordinates. In field-aligned
coordinates, a scalar function is provided with the following $3$
properties.

$1$, periodic condition in $\alpha$ direction 
\begin{equation}
f\left(\psi,\theta,\alpha\right)=\sum_{n}f_{n}\left(\psi,\theta\right)\me^{\mi n\alpha}.\label{eq:subscript_n}
\end{equation}
Thus, the partial differential operator in $\alpha$ direction $\partial_{\alpha}$
is converted to the algebraic operator $\mi n$.

$2$, field-aligned periodic condition in $\theta$ direction 
\begin{equation}
f\left(\psi,\theta\pm2\pi,\alpha\right)=f\left(\psi,\theta,\alpha\mp2q\left(\psi\right)\pi\right),
\end{equation}
in the form of Fourier components
\begin{equation}
f_{n}\left(\psi_{i},\theta_{j}\pm2\pi\right)=f_{n}\left(\psi_{i},\theta_{j}\right)\me^{\mp2\mi nq_{i}\pi},
\end{equation}
which ensures that the central difference formula can be used at the
boundary of $\theta$.

$3$, single valued condition at the magnetic axis. In magnetic flux
coordinates, a scalar function $f_{f}\left(\bm{X}_{f}\right)$ satisfies
\begin{equation}
\partial_{\theta^{*}}f_{f}\biggl|_{\psi^{*}=0}=0.
\end{equation}
In field-aligned coordinates, $f\left(\psi,\theta,\alpha\right)=f_{f}\left(\psi^{*},\theta^{*},\zeta^{*}\right)$.
By using Eqs.~(\ref{eq:flux2field1})-(\ref{eq:flux2field3}), we
can obtain 
\begin{equation}
f\left(\psi_{0},\theta,\alpha\right)=f\left(\psi_{0},\theta_{0},\alpha+q_{0}\left(\theta_{0}-\theta\right)\right),
\end{equation}
in the form of Fourier components 
\begin{equation}
f_{n}\left(\psi_{0},\theta_{j}\right)=f_{n}\left(\psi_{0},\theta_{0}\right)\me^{\mi nq_{0}\left(\theta_{0}-\theta_{j}\right)}.
\end{equation}
The partial differential operator in $\theta$ direction $\partial_{\theta}$
are converted to the algebraic operator $-inq_{0}$.

The boundary condition in radial direction is $\phi\left(\psi_{N_{\psi}-1}\right)=0$.
For each toroidal mode number $n$, by considering that $N_{\theta}-1$
equations are obtained with the single valued condition at the magnetic
axis, we need another $N_{\theta}\times\left(N_{\psi}-2\right)+1$
equations to solve the perturbed field. By integrating both sides
of the Eq.~(\ref{eq:QN}) with $\int\md\psi\md\theta J_{\bm{X}}$
and discretizing it numerically, we can obtain 
\begin{align}
 & A_{n,\fV}\Phi_{n,\fV}=R_{n,\fV}\left[\rho_{n}\right]_{\fV},\label{eq:AE_QN}\\
 & A_{n,\fV}\equiv D_{n,\fV}-P_{n,\fV}+Z_{n,\fV},
\end{align}
where $\fV$ is the integral domain, $D_{n,\fV}\Phi_{n,\fV}$, $P_{n,\fV}\Phi_{n,\fV}$,
$Z_{n,\fV}\Phi_{n,\fV}$ represent contributions of $\int_{\fV_{i,j}}\md\psi\md\theta J_{\bm{X}}\nabla\cdot\left(c_{0}\nabla_{\perp}\phi_{n}\right)$,
$\int_{\fV_{i,j}}\md\psi\md\theta J_{\bm{X}}c_{1}\phi_{n}$, $\int_{\fV_{i,j}}\md\psi\md\theta J_{\bm{X}}c_{1}\langle\phi\rangle_{FA}$,
respectively; the subscript $n$ is understood in the way similar
to Eq.~(\ref{eq:subscript_n}). If $n\ne0$, then $Z_{n,\fV}=0$.
In numerical, differential is discretized by using the central difference
formula, the surface integral of the electric field flux is discretized
by using the composite midpoint rule, the integral of the ion density
is discretized by using the composite trapezoidal rule.

In the non-magnetic axis domain, $i=1,2,\cdots,N_{\psi}-2$, the integral
region is $\fV_{i,j}=\left[\psi_{i}-\frac{1}{2}\Delta_{\psi},\psi_{i}+\frac{1}{2}\Delta_{\psi}\right]\times\left[\theta_{j}-\frac{1}{2}\Delta_{\theta},\theta_{j}+\frac{1}{2}\Delta_{\theta}\right]$.
We can obtain that
\begin{align}
D_{n,\fV_{i,j}}\Phi_{n,\fV_{i,j}} & =\sum_{i'=i-1}^{i+1}\sum_{j'=j-1}^{j+1}d_{n,\fV_{i,j}}^{i',j'}\phi_{n}^{i',j'},\\
P_{n,\fV_{i,j}}\Phi_{n,\fV_{i,j}} & =p_{n,\fV_{i,j}}^{i,j}\phi_{n}^{i,j},\\
Z_{n,\fV_{i,j}}\Phi_{n,\fV_{i,j}} & =\begin{cases}
0 & n\ne0,\\
\sum_{i'=i-1}^{i+1}\sum_{j'=0}^{N_{\theta}-1}z_{0,\fV_{i,j}}^{i',j'}\phi_{0}^{i',j'} & n=0,
\end{cases}\label{eq:ZPhi}\\
R_{n,\fV_{i,j}}\left[\rho_{n}\right]_{\fV_{i,j}} & =\sum_{i'=i-1}^{i+1}r_{n,\fV_{i,j}}^{i',j}\rho_{n}^{i',j}.
\end{align}
For example, we compute the $z_{0,\fV_{i,j}}$ used in Eq.~(\ref{eq:ZPhi}),
which represents the contribution of $\int_{\fV_{i,j}}\md\psi\md\theta J_{\bm{X}}c_{1}\langle\phi\rangle_{FA}$.
If $i>1$, by using Eq.~(\ref{eq:FA}), we can obtain that
\begin{align}
\int_{\fV_{i,j}}\md\psi\md\theta J_{\bm{X}}c_{1}\langle\phi\rangle_{FA}= & \frac{1}{8}\Delta_{\psi}\Delta_{\theta}J_{\bm{X}}^{i-1,j}c_{1}^{i-1,j}\frac{\sum_{j'=0}^{N_{\theta}-1}J_{\bm{X}}^{i-1,j'}\phi_{0}^{i-1,j'}}{\sum_{j''=0}^{N_{\theta}-1}J_{\bm{X}}^{i-1,j''}}\nonumber \\
 & +\frac{6}{8}\Delta_{\psi}\Delta_{\theta}J_{\bm{X}}^{i,j}c_{1}^{i,j}\frac{\sum_{j'=0}^{N_{\theta}-1}J_{\bm{X}}^{i,j'}\phi_{0}^{i,j'}}{\sum_{j''=0}^{N_{\theta}-1}J_{\bm{X}}^{i,j''}}\nonumber \\
 & +\frac{1}{8}\Delta_{\psi}\Delta_{\theta}J_{\bm{X}}^{i+1,j}c_{1}^{i+1,j}\frac{\sum_{j'=0}^{N_{\theta}-1}J_{\bm{X}}^{i+1,j'}\phi_{0}^{i+1,j'}}{\sum_{j''=0}^{N_{\theta}-1}J_{\bm{X}}^{i+1,j''}},
\end{align}
thus,
\begin{equation}
z_{0,\fV_{i,j}}^{i',j'}=\begin{cases}
\frac{1}{8}\Delta_{\psi}\Delta_{\theta}J_{\bm{X}}^{i-1,j}c_{1}^{i-1,j}\frac{J_{\bm{X}}^{i-1,j'}}{\sum_{j''=0}^{N_{\theta}-1}J_{\bm{X}}^{i-1,j''}} & i'=i-1,\\
\frac{6}{8}\Delta_{\psi}\Delta_{\theta}J_{\bm{X}}^{i,j}c_{1}^{i,j}\frac{J_{\bm{X}}^{i,j'}}{\sum_{j''=0}^{N_{\theta}-1}J_{\bm{X}}^{i,j''}} & i'=i,\\
\frac{1}{8}\Delta_{\psi}\Delta_{\theta}J_{\bm{X}}^{i+1,j}c_{1}^{i+1,j}\frac{J_{\bm{X}}^{i+1,j'}}{\sum_{j''=0}^{N_{\theta}-1}J_{\bm{X}}^{i+1,j''}} & i'=i+1.
\end{cases}
\end{equation}
Else if $i=1$, the fomulation of the magnetic surface average at
the magnetic axis need to be used. It is reduced to
\begin{equation}
\left\langle f\right\rangle _{FA}|_{\psi=0}=\frac{1}{4\pi^{2}}\int_{0}^{2\pi}\md\alpha\int_{0}^{2\pi}\md\theta f=f_{0}|_{\psi=0},
\end{equation}
which can be used in simulation even if the space Jacobian is equal
to zero. Note that $J_{\bm{X}}$ and $c_{1}$ is independent of $\theta$
at the axis, it is not difficult to obtain that
\begin{equation}
z_{0,\fV_{1,j}}^{i',j'}=\begin{cases}
\frac{1}{8}\Delta_{\psi}\Delta_{\theta}J_{\bm{X}}^{0,0}c_{1}^{0,0} & i'=0,\\
\frac{6}{8}\Delta_{\psi}\Delta_{\theta}J_{\bm{X}}^{1,j}c_{1}^{1,j}\frac{J_{\bm{X}}^{1,j'}}{\sum_{j''=0}^{N_{\theta}-1}J_{\bm{X}}^{1,j''}} & i'=1,\\
\frac{1}{8}\Delta_{\psi}\Delta_{\theta}J_{\bm{X}}^{2,j}c_{1}^{2,j}\frac{J_{\bm{X}}^{2,j'}}{\sum_{j''=0}^{N_{\theta}-1}J_{\bm{X}}^{2,j''}} & i'=2.
\end{cases}
\end{equation}
It is worth pointing out that if $j=0$ or $N_{\theta}-1$, then $\phi_{n}^{i,-1}$
or $\phi_{n}^{i,N_{\theta}}$ will appear in the computation. By using
the field-aligned periodic condition, we can obtain $\phi_{n}^{i,-1}=\phi_{n}^{i,N_{\theta}-1}\me^{2\mi nq_{i}\pi}$
and $\phi_{n}^{i,N_{\theta}}=\phi_{n}^{i,0}\me^{-2\mi nq_{i}\pi}$,
and absorb $\me^{\pm2\mi nq_{i}\pi}$ into $z_{0,\fV_{i,j}}^{i',j'}$.
The discretized equation can always be written in the form of Eq.~(\ref{eq:AE_QN}).

At the magnetic axis, $i=0$, the integral domain is $\fV_{A}=\left[0,\frac{1}{2}\Delta_{\psi}\right]\times\left[-\pi-\frac{1}{2}\Delta_{\theta},\pi-\frac{1}{2}\Delta_{\theta}\right]$.
We have
\begin{align}
D_{n,\fV_{A}}\Phi_{n,\fV_{A}} & =d_{n,\fV_{A}}^{0,0}\phi_{n}^{0,0}+\sum_{j=0}^{N_{\theta}-1}d_{n,\fV_{A}}^{1,j}\phi_{n}^{1,j},\label{eq:DPhi}\\
P_{n,\fV_{A}}\Phi_{n,\fV_{A}} & =p_{n,\fV_{A}}^{0,0}\phi_{n}^{0,0}+\sum_{j'=0}^{N_{\theta}-1}p_{n,\fV_{A}}^{1,j'}\phi_{0}^{1,j'},\\
Z_{n,\fV_{A}}\Phi_{n,\fV_{A}} & =\begin{cases}
0 & n\ne0,\\
z_{0,\fV_{A}}\phi_{0}^{0,0}+\sum_{j'=0}^{N_{\theta}-1}z_{0,\fV_{A}}^{1,j'}\phi_{0}^{1,j'} & n=0,
\end{cases}\\
R_{n,\fV_{A}}\left[\rho_{n}\right]_{\fV_{A}} & =r_{n,\fV_{A}}^{0,0}\rho_{n}^{0,0}+\sum_{j'=0}^{N_{\theta}-1}r_{n,\fV_{A}}^{1,j'}\rho_{n}^{1,j'}.
\end{align}
For example, we compute the $d_{n,\fV_{A}}$ used in Eq.~(\ref{eq:DPhi}),
which represent the contribution of $\int_{\fV_{i,j}}\md\psi\md\theta J_{\bm{X}}\nabla\cdot\left(c_{0}\nabla_{\perp}\phi_{n}\right)$.
\begin{equation}
\int_{\fV_{A}}\md\psi\md\theta J_{s}\nabla\cdot\left(c_{0}\nabla_{\perp}\phi_{n}\right)=I_{n,\fV_{A}}^{\psi}+I_{n,\fV_{A}}^{\theta}+I_{n,\fV_{A}}^{\alpha},
\end{equation}
 with
\begin{align}
I_{n,\fV_{A}}^{\psi}\equiv & \int\md\psi\md\theta\partial_{\psi}\left(C_{\psi\psi}\partial_{\psi}\phi_{n}+C_{\psi\theta}\partial_{\theta}\phi_{n}+\mi nC_{\psi\alpha}\phi_{n}\right)\nonumber \\
= & \int_{-\pi-\frac{1}{2}\Delta_{\theta}}^{\pi-\frac{1}{2}\Delta_{\theta}}\md\theta\left(C_{\psi\psi}\partial_{\psi}\phi_{n}+C_{\psi\theta}\partial_{\theta}\phi_{n}+\mi nC_{\psi\alpha}\phi_{n}\right)|_{\psi=\frac{1}{2}\Delta_{\psi}},\\
I_{n,\fV_{A}}^{\theta}\equiv & \int_{\fV_{A}}\md\psi\md\theta\partial_{\theta}\left(C_{\theta\psi}\partial_{\psi}\phi_{n}+C_{\theta\theta}\partial_{\theta}\phi_{n}+\mi nC_{\theta\alpha}\phi_{n}\right)\nonumber \\
= & \int_{0}^{\frac{1}{2}\Delta_{\psi}}\md\psi\left(C_{\theta\psi}\partial_{\psi}\phi_{n}+C_{\theta\theta}\partial_{\theta}\phi_{n}+\mi nC_{\theta\alpha}\phi_{n}\right)|_{\theta=\pi-\frac{1}{2}\Delta_{\psi}}\nonumber \\
 & -\int_{0}^{\frac{1}{2}\Delta_{\psi}}\md\psi\left(C_{\theta\psi}\partial_{\psi}\phi_{n}+C_{\theta\theta}\partial_{\theta}\phi_{n}+\mi nC_{\theta\alpha}\phi_{n}\right)|_{\theta=-\pi-\frac{1}{2}\Delta_{\psi}},\\
I_{n,\fV_{A}}^{\alpha}\equiv & \int_{\fV_{A}}\md\psi\md\theta\mi n\left(C_{\alpha\psi}\partial_{\psi}\phi_{n}+C_{\alpha\theta}\partial_{\theta}\phi_{n}+C_{\alpha\alpha}\mi n\phi_{n}\right),
\end{align}
and
\begin{equation}
\begin{aligned}C_{X_{l}^{a}X_{l}^{b}} & \left(\psi,\theta\right)=\begin{cases}
J_{s}c_{0}\nabla\theta\cdot\nabla\theta-\frac{c_{0}}{J_{\bm{X}}B^{2}}, & X_{l}^{a}=X_{l}^{b}=\theta\\
J_{s}c_{0}\nabla X_{l}^{a}\cdot\nabla X_{l}^{b}, & other
\end{cases}.\end{aligned}
\end{equation}
The natural boundary condition 
\begin{equation}
\left(C_{\psi\psi}\partial_{\psi}\phi_{n}+C_{\psi\theta}\partial_{\theta}\phi_{n}+\mi nC_{\psi\alpha}\phi_{n}\right)_{\psi=0}=0,
\end{equation}
is always satisfied, which makes the artificial boundary condition
unnecessary. If $\psi$ is chosen as the radial coordinate $x$, $\nabla x\cdot\nabla x\partial_{x}\phi=\nabla x\cdot\nabla\theta\partial_{\theta}\phi=\nabla x\cdot\nabla\alpha\partial_{\alpha}\phi=0$;
If $\sqrt{\psi}$ or $r$ is chosen as $x$, $J_{x,\theta,\alpha}\partial_{x}\phi=J_{x,\theta,\alpha}\partial_{\theta}\phi=J_{x,\theta,\alpha}\partial_{\alpha}\phi=0$.
It can be obtained that 
\begin{equation}
\begin{aligned}d_{n,\fV_{A}}^{0,0}= & \left[-\frac{1}{2}C_{\psi\theta}^{1/4,N_{\theta}-1/2}-\frac{3}{8}\mi n\Delta_{\psi}\left(q_{0}C_{\theta\theta}^{1/4,N_{\theta}-1/2}-C_{\theta\alpha}^{1/4,N_{\theta}-1/2}\right)\right]\me^{-\mi nq_{0}(2\pi-\Delta_{\theta}/2)}\\
 & +\left[\frac{1}{2}C_{\psi\theta}^{1/4,-1/2}+\frac{3}{8}\mi n\Delta_{\psi}\left(q_{0}C_{\theta\theta}^{1/4,-1/2}-C_{\theta\alpha}^{1/4,-1/2}\right)\right]\me^{\mi nq_{0}\Delta_{\theta}/2}\\
 & -\sum_{j=0}^{N_{\theta}-1}\biggl[h_{\psi}\Delta_{\theta}C_{\psi\psi}^{1/2,j}+\frac{1}{2}\mi n\Delta_{\theta}\left(q_{0}C_{\psi\theta}^{1/2,j}-C_{\psi\alpha}^{1/2,j}+c_{\psi\alpha}^{1/4,j}\right)\\
 & -\frac{3}{8}n^{2}\Delta_{\psi}\Delta_{\theta}\left(q_{0}C_{\theta\alpha}^{1/4,j}-C_{\alpha\alpha}^{1/4,j}\right)\biggr]\me^{\mi nq_{0}(\theta_{0}-\theta_{j})},
\end{aligned}
\end{equation}
and
\begin{align}
d_{n,\fV_{A}}^{1,j}= & \begin{cases}
h_{\psi}\Delta_{\theta}C_{\psi\psi}^{1/2,0}+\frac{1}{2}\mi n\Delta_{\theta}C_{\psi\alpha}^{1/2,0}\\
+\frac{1}{4}\left(C_{\psi\theta}^{1/2,N_{\theta}-1}\me^{-\mi n2q_{1}\pi}-C_{\psi\theta}^{1/2,1}\right)+\frac{1}{2}\mi n\Delta_{\theta}C_{\psi\alpha}^{1/4,0}\\
+\frac{1}{16}\mi n\Delta_{\psi}\left(C_{\theta\alpha}^{1/4,N_{\theta}-1}\me^{-\mi n2q_{1}\pi}-C_{\theta\alpha}^{1/4,1}\right)-\frac{1}{8}n^{2}\Delta_{\psi}\Delta_{\theta}C_{\alpha\alpha}^{1/4,0}\\
+\left(\frac{1}{4}C_{\psi\theta}^{1/4,N_{\theta}-1/2}+\frac{1}{8}\Delta_{\psi}h_{\theta}C_{\theta\theta}^{1/4,N_{\theta}-1/2}+\frac{1}{16}\mi n\Delta_{\psi}C_{\theta\alpha}^{1/4,N_{\theta}-1/2}\right)\me^{-\mi n2q_{1}\pi}\\
-\left(\frac{1}{4}C_{\psi\theta}^{1/4,-1/2}+\frac{1}{8}\Delta_{\psi}h_{\theta}c_{\theta\theta}^{1/4,-1/2}+\frac{1}{16}\mi n\Delta_{\psi}C_{\theta\alpha}^{1/4,-1/2}\right), & j=0\\
\\
h_{\psi}\Delta_{\theta}C_{\psi\psi}^{1/2,N_{\theta}-1}+\frac{1}{2}\mi n\Delta_{\theta}C_{\psi\alpha}^{1/2,N_{\theta}-1}\\
+\frac{1}{4}\left(C_{\psi\theta}^{1/2,N_{\theta}-2}-C_{\psi\theta}^{1/2,0}\me^{\mi n2q_{1}\pi}\right)+\frac{1}{2}\mi n\Delta_{\theta}C_{\psi\alpha}^{1/4,N_{\theta}-1}\\
+\frac{1}{16}\mi n\Delta_{\psi}\left(C_{\theta\alpha}^{1/4,N_{\theta}-2}-C_{\theta\alpha}^{1/4,0}\me^{\mi n2q_{1}\pi}\right)-\frac{1}{8}n^{2}\Delta_{\psi}\Delta_{\theta}C_{\alpha\alpha}^{1/4,N_{\theta}-1}\\
+\left(\frac{1}{4}C_{\psi\theta}^{1/4,N_{\theta}-1/2}-\frac{1}{8}\Delta_{\psi}h_{\theta}C_{\theta\theta}^{1/4,N_{\theta}-1/2}+\frac{1}{16}\mi n\Delta_{\psi}C_{\theta\alpha}^{1/4,N_{\theta}-1/2}\right)\\
-\left(\frac{1}{4}C_{\psi\theta}^{1/4,-1/2}-\frac{1}{8}\Delta_{\psi}h_{\theta}C_{\theta\theta}^{1/4,-1/2}+\frac{1}{16}\mi n\Delta_{\psi}C_{\theta\alpha}^{1/4,-1/2}\right)\me^{\mi n2q_{1}\pi}, & j=N_{\theta}-1\\
\\
h_{\psi}\Delta_{\theta}C_{\psi\psi}^{1/2,j}+\frac{1}{2}\mi n\Delta_{\theta}\left(c_{\psi\alpha}^{1/2,j}+C_{\psi\alpha}^{1/4,j}\right)+\frac{1}{4}(C_{\psi\theta}^{1/2,j-1}-C_{\psi\theta}^{1/2,j+1})\\
+\frac{1}{16}\mi n\Delta_{\psi}\left(C_{\theta\alpha}^{1/4,j-1}-C_{\theta\alpha}^{1/4,j+1}\right)-\frac{1}{8}n^{2}\Delta_{\psi}\Delta_{\theta}C_{\alpha\alpha}^{1/4,j}. & other
\end{cases}
\end{align}
where $C_{X_{l}^{a}X_{l}^{b}}^{i,j}\equiv C_{X_{l}^{a}X_{l}^{b}}\left(i\Delta_{\psi},\theta_{0}+j\Delta_{\theta}\right)$.
Similarly, we can obtain elements $p_{n,\fV_{A}}^{i,j}$, $z_{n,\fV_{A}}^{i,j}$,
$r_{n,\fV_{A}}^{i,j}$ of matrix $P_{n,\fV_{A}}$, $Z_{n,\fV_{A}}$,
$R_{n,\fV_{A}}$, respectively. 

Thus, $1$ equation is obtained at the magnetic axis domain. Combined
with $N_{\theta}\times\left(N_{\psi}-2\right)$ equations obtained
in the non-magnetic axis domain and another $N_{\theta}-1$ equations
obtained by the single valued condition at the magnetic axis, we have
got Eq.~(\ref{eq:AE_QN}), $N_{\theta}\times\left(N_{\psi}-1\right)$
equations in total, for solving the perturbed electrostatic potential.

To reduce the computational cost, we organize the field matrix as
follow
\begin{align}
 & \begin{aligned}\Phi_{n}\end{aligned}
=\left[\begin{array}{ccccc}
\Phi_{n,0}, & \cdots, & \Phi_{n,i}, & \cdots, & \Phi_{n,N_{\psi}-1}\end{array}\right]^{T},\\
 & \Phi_{n,i}=\left[\phi_{n}^{i,0},\cdots,\phi_{n}^{i,j},\cdots,\phi_{n}^{i,N_{\theta}-1}\right],
\end{align}
and $\left[\rho_{n}\right]$ is also organized as $\Phi_{n}$. Thus,
the coefficient matrix $\left[A_{n}\right]$ is composed of block
matrix $\left[A_{n}^{i,i'}\right]$
\begin{align}
 & A_{n}=\left[A_{n}^{i,i'}\right],\\
 & A_{n}^{i,i'}=\left[a_{n,i,i'}^{j,j'}\right],\\
 & a_{n,i,i'}^{j,j'}=\begin{cases}
d_{n,\fV_{i,j}}^{i',j'}-p_{n,\fV_{i,j}}^{i',j'}+z_{n,\fV_{i,j}}^{i',j'} & i\ne0,\\
d_{n,\fV_{A}}^{i',j'}-p_{n,\fV_{A}}^{i',j'}+z_{n,\fV_{A}}^{i',j'} & i=0,j=0,\\
\delta_{ii'}\me^{\mi nq_{0}\left(\theta_{0}-\theta_{j}\right)} & i=0,j\ne0,j'=0,\\
-\delta_{ii'}\delta_{jj'} & i=0,j\ne0,j'\ne0,
\end{cases}
\end{align}
and $a_{n,i,i'}^{j,j'}$ represents the matrix element of $A_{n}^{i,i'}$.
We can easily find that $A_{n}^{i,i'}=0$ if $\left|i'-i\right|>1$,
$A_{n}$ is a block tridiagonal matrix. The algebraic equations can
be solved by using the tridiagonal matrix algorithm.

\section{Numerical filter}

\label{Sec:6}

On the one hand, because the inner damping buffer region is not used,
the filter in the radial direction is replaced by the low-pass filter.
On the other hand, an additional condition for truncating the shortwave
in poloidal direction is used,
\begin{equation}
k_{\theta}\rho_{s}=\frac{|m|}{r}\frac{\sqrt{m_{i}T_{i}}}{e_{i}B}<\left[k_{\theta}\rho\right]_{max}.
\end{equation}
Thus, we have 
\begin{equation}
|m|<m_{max}\equiv\min\left\{ r\frac{e_{i}B}{\sqrt{m_{i}T_{i}}}\left[k_{\theta}\rho\right]_{max},\frac{\tilde{N}_{\theta}}{3}\right\} ,
\end{equation}
where $\tilde{N}_{\theta^{*}}$ represents the total number of the
poloidal grid point in the $\theta^{*}$ direction. It is necessary
that $N_{\theta^{*}}\ll N_{\theta}$ to describe poloidal mode structures.
By considering the parallel wavelength truncated condition $-\frac{N_{\theta}}{3}<m-nq\left(\psi\right)<\frac{N_{\theta}}{3}$,
we have
\begin{equation}
|n|<n_{max}\equiv\min\left\{ \frac{1}{q}\left(m_{max}+\frac{N_{\theta}}{3}\right),\frac{N_{\alpha}}{3}\right\} .
\end{equation}
Finally, the filter conditions of $m$ and $n$ are 
\begin{align}
n & \in\left(-n_{max},n_{max}\right),\\
m & \in\left(-m_{max},m_{max}\right)\text{\ensuremath{\bigcap}}\left(nq-\frac{N_{\theta}}{3},nq+\frac{N_{\theta}}{3}\right).
\end{align}
In the ITG turbulence, the typical value of $k_{\perp}\rho_{i}$ is
about $0.3$, thus we take $\left[k_{\theta}\rho\right]_{max}=3$
and $\tilde{N}_{\theta}=512$.

\begin{figure}
\centering{}\includegraphics[width=0.45\textwidth]{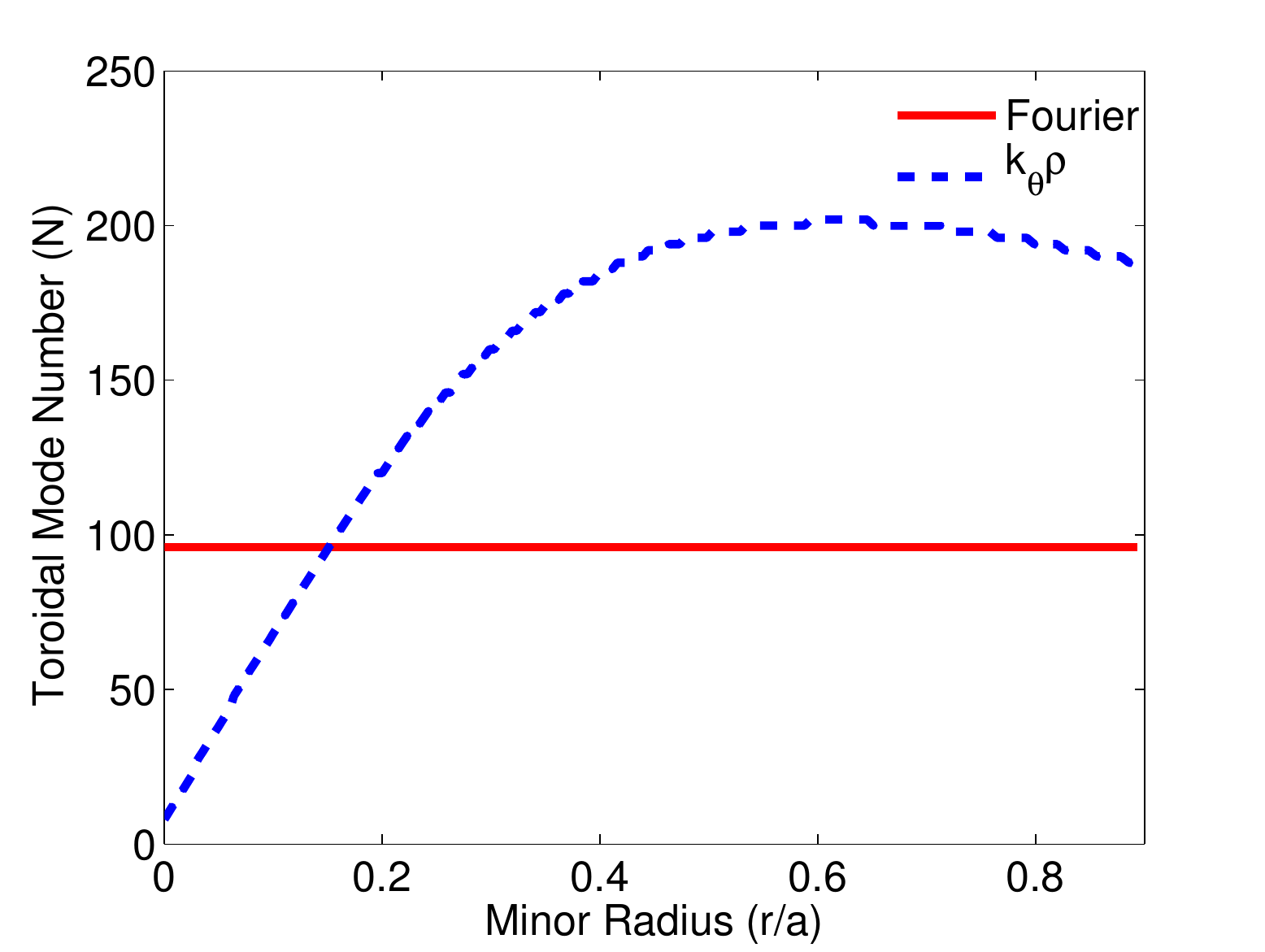}\caption{\label{fig:Truncation}Relation between the truncation of the toroidal
mode number and the radial positon. Red solid line: The truncated
condition decided by the parallel and the poloidal wavelength, $\frac{1}{q}\left(m_{max}+\frac{N_{\theta}}{3}\right)$.
Blue dash line: The truncated condition decided by the toroidal wavelength.
Read solid line is the truncated condition, $\frac{N_{\alpha}}{3}$.
The applied truncated condition in simulation is determined by the
smaller one of above $2$ conditions.}
\end{figure}
As is shown in Fig.~(\ref{fig:Truncation}), the untruncated toroidal
mode number in the ITG simulation is reduced with the radius near
the magnetic axis. 

\section{Simulation results}

\label{Sec:7}

In this section, nonlinear simulation results of the cyclone base
test are shown. The parameters are chosen as those in Ref.~\citep{2010-Lapillonne-Benchmark}
to compare with the results computed by GENE, ORB5. $q$ profile is
\begin{equation}
q\left(r\right)=0.86-0.16\frac{r}{a}+2.52\left(\frac{r}{a}\right)^{2}.
\end{equation}
$q\left(r_{0}\right)=1.41$, magnetic share $\hat{s}\left(r_{0}\right)\equiv\frac{r}{q}\frac{\md q}{\md r}\left(r_{0}\right)=0.84$
with $r_{0}=0.5a$. The initial ion temperature and density profile
are
\begin{equation}
\hat{A}\left(r\right)=\frac{A\left(r\right)}{A\left(r_{0}\right)}=\exp\left[-\kappa_{A}\frac{a}{R_{0}}\Delta_{A}\tanh\left(\frac{r-r_{0}}{a}\right)\right],
\end{equation}
where $A$ can be chosen as either $T_{i}$ or $n_{i}$, and $T_{i}\left(r_{0}\right)=1.97\mkeV$,
$n\left(r_{0}\right)=10^{19}\mm^{-3}$, $\Delta_{A}=0.30$, $\kappa_{n}\equiv R_{0}/L_{n}=2.23$,
$\kappa_{T_{i}}=R_{0}/L_{T_{i}}=6.96$. $L_{n}$ and $L_{T_{i}}$
represents the scale length of density and ion temperature, respectively.
The pure deuterium ion and the adiabatic electron are adopted. The
ratio of the ion Larmor radius and the minor radius is $\rho^{*}\equiv\rho_{s}/a=1/179$
with $\rho_{s}=c_{s}/\Omega_{i}$, $c_{s}=\sqrt{T_{i0}/m_{i}}$, $\Omega_{i}=e_{i}B_{0}/m_{i}$.
The radial simulation domain including the magnetic axis is about
$160\rho_{s}$, the grid resolution of this simulation is taken as
$N_{\psi}\times N_{\theta}\times N_{\alpha}\times N_{V_{\parallel}}\times N_{\mu}=189\times16\times141\times64\times16$.
Note that if $\psi$ is chosen as the radial coordinate, the radial
resolution is coarse near the magnetic axis. The simulation with $\sqrt{\psi}$
being the radial coordinate is also computed as a comparison. In the
rest of this section, NLT1, NLT2, NLT3 represent simulations by using
the version of NLT without magnetic axis, with magnetic axis and $\psi$
being the radial coordinate, with magnetic axis and $\sqrt{\psi}$
being the radial coordinate, respectively. 

\begin{figure}
\centering{}\includegraphics[width=0.45\textwidth]{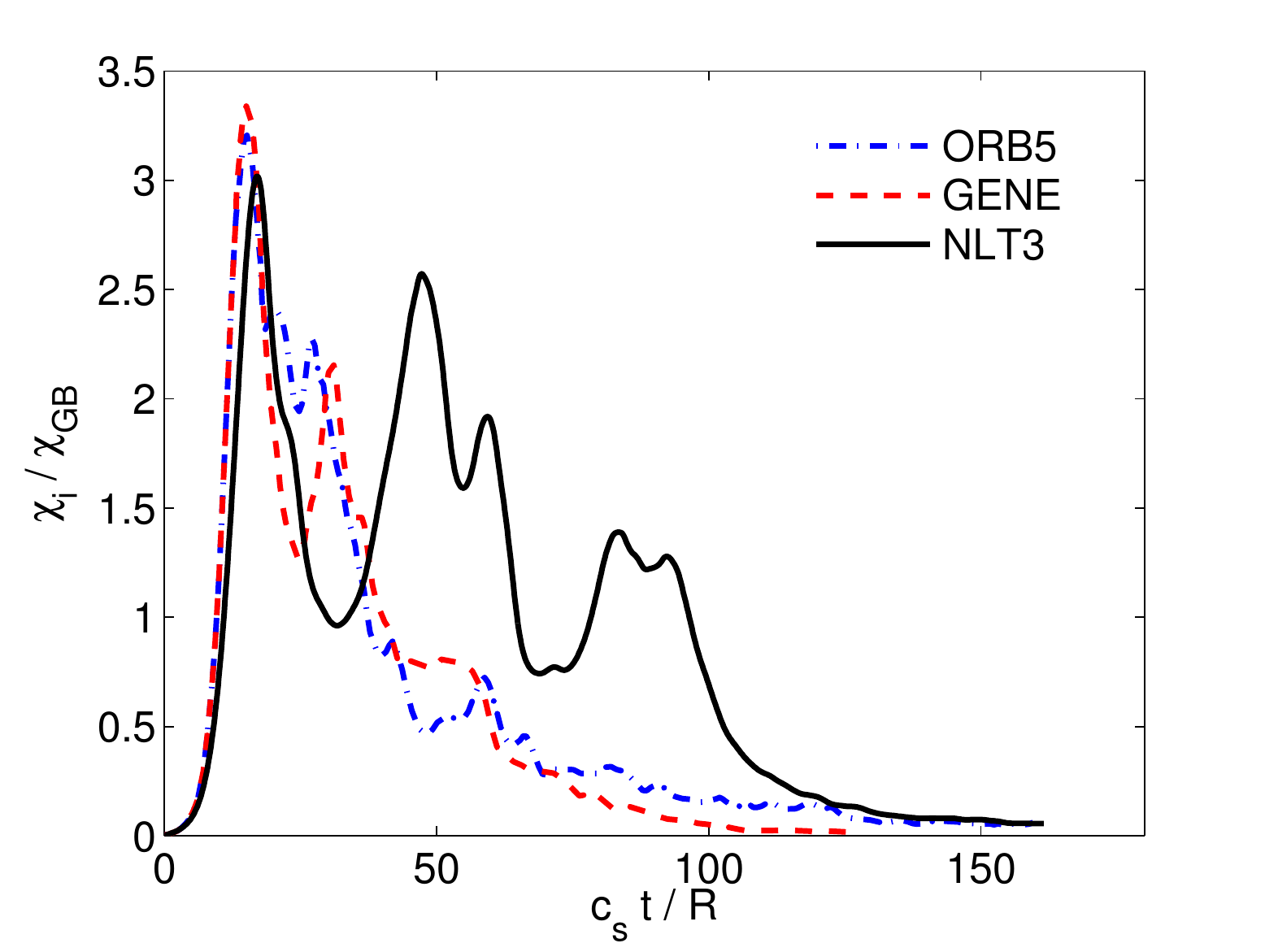}\includegraphics[clip,width=0.45\textwidth]{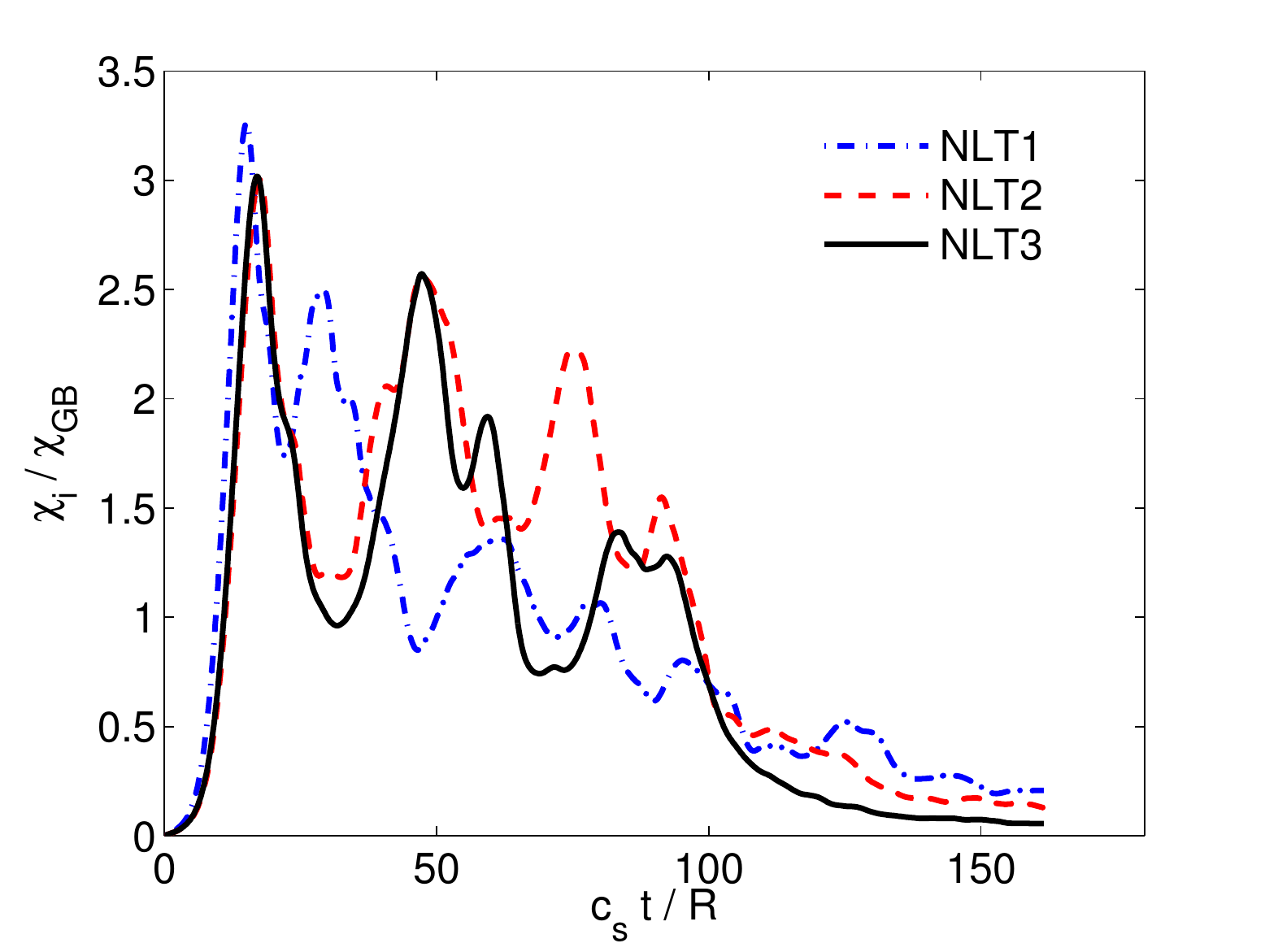}\caption{\label{fig:kai}Time evolutions of the ion heat diffusivity $\chi_{i}/\chi_{GB}$,
with $\chi_{GB}\equiv\frac{\rho_{i}^{2}c_{s}}{a}$.}
\end{figure}

\begin{figure}
\centering{}\includegraphics[width=0.45\textwidth]{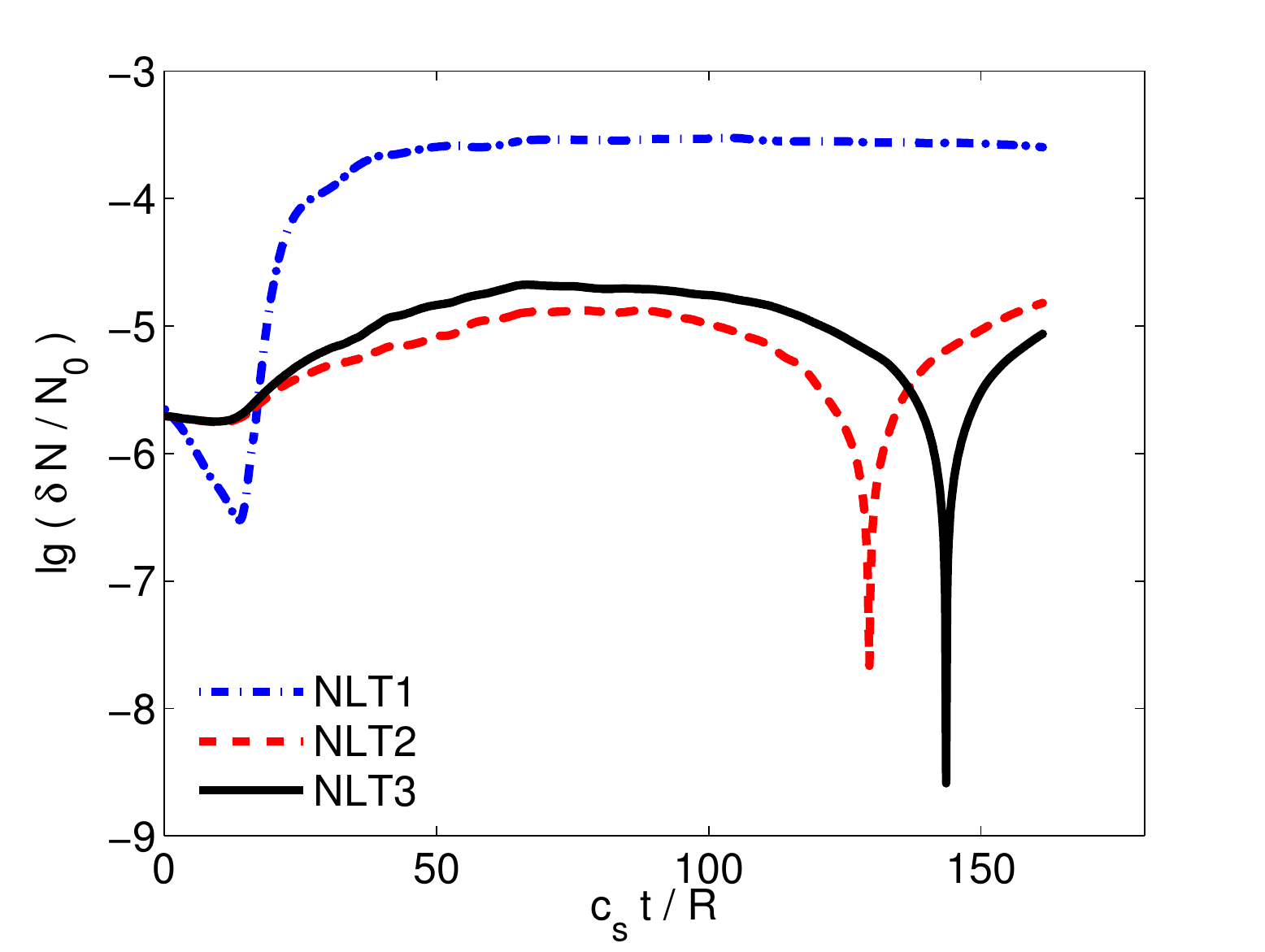}\caption{\label{fig:Number}Time evolutions of the perturbed ion gyrocenter
number, with $\delta N=\int\md^{6}Z\delta F$, $N_{0}=\int\md^{6}ZF_{0}$.}
\end{figure}

\begin{figure}
\centering{}\includegraphics[width=0.45\textwidth]{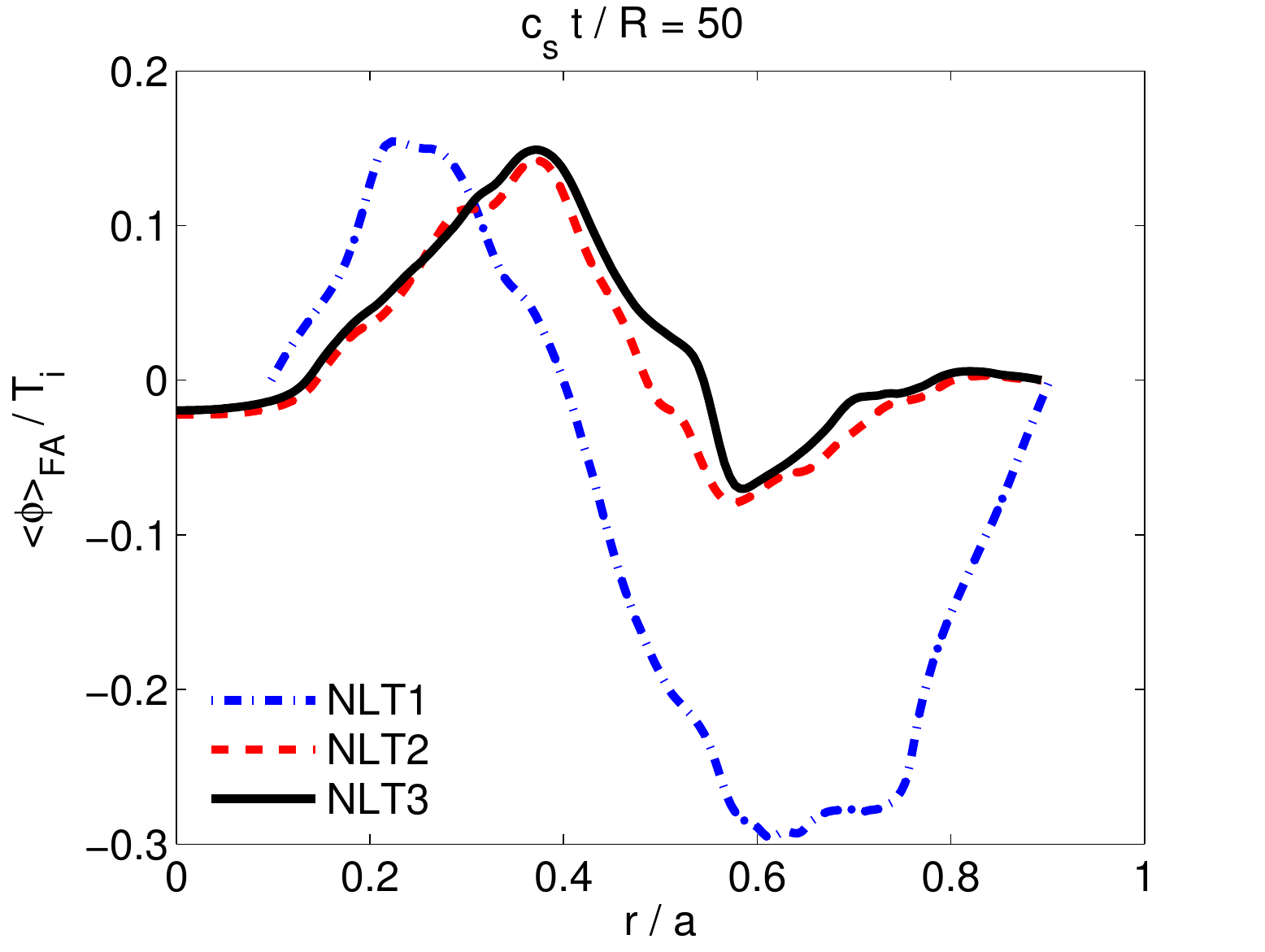}\includegraphics[width=0.45\textwidth]{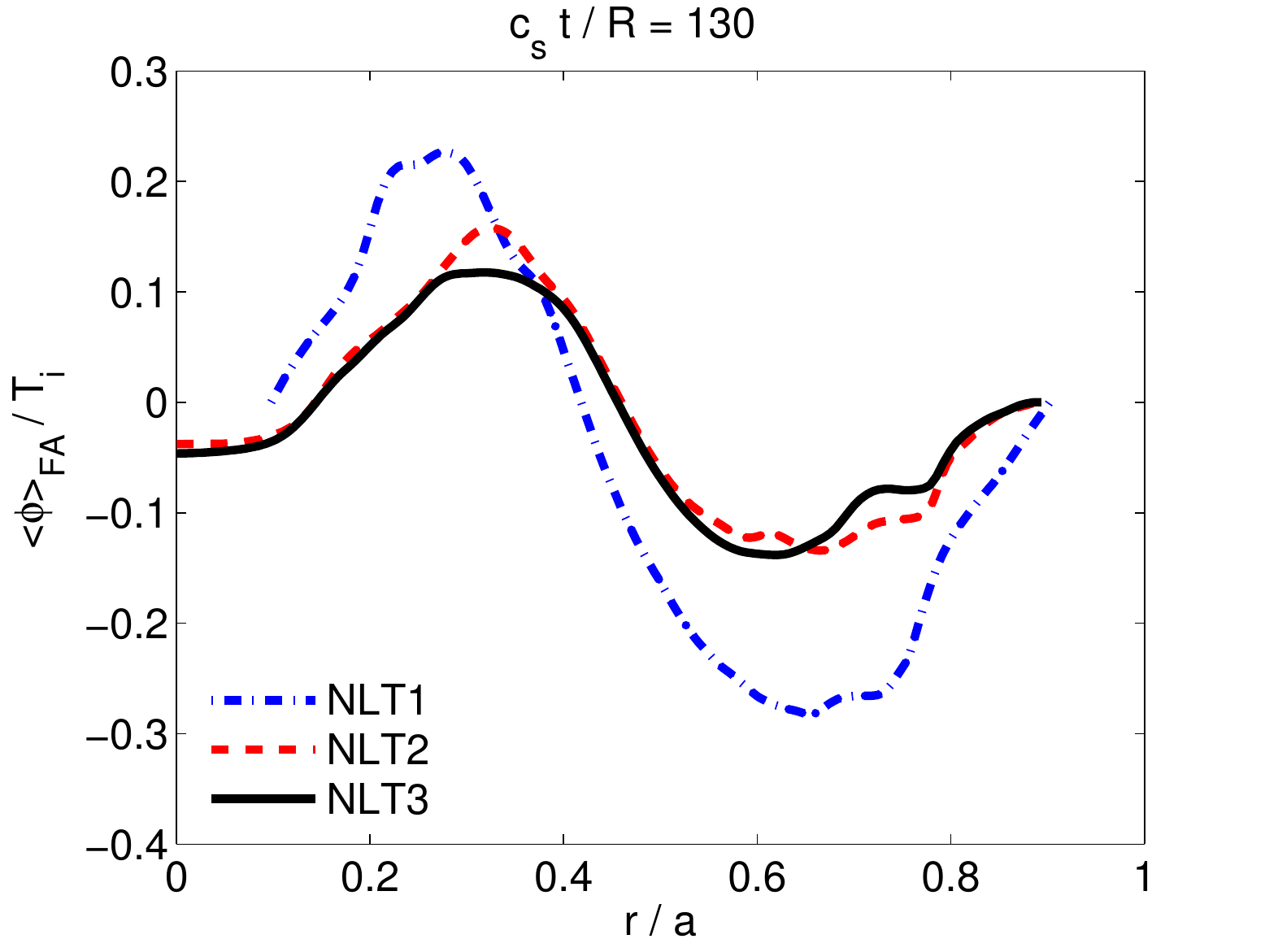}\caption{\label{fig:phiZ}Zonal field $\left\langle \phi\right\rangle _{FA}$.
Left and right figures show the zonal field at $\frac{c_{s}t}{R}=50$
and $\frac{c_{s}t}{R}=130$, respectively.}
\end{figure}

\begin{figure}
\begin{centering}
\includegraphics[width=0.45\textwidth]{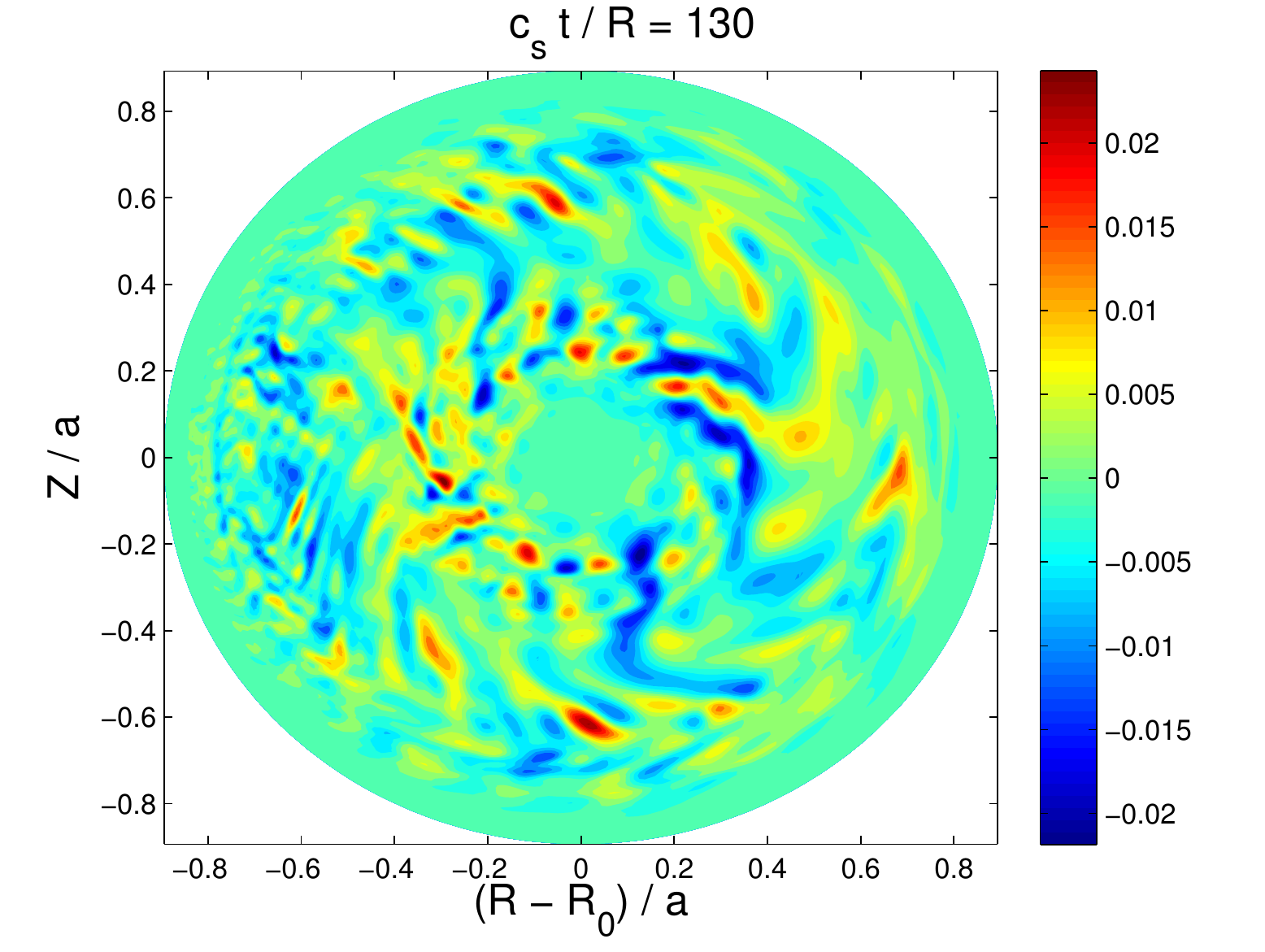}\includegraphics[width=0.45\textwidth]{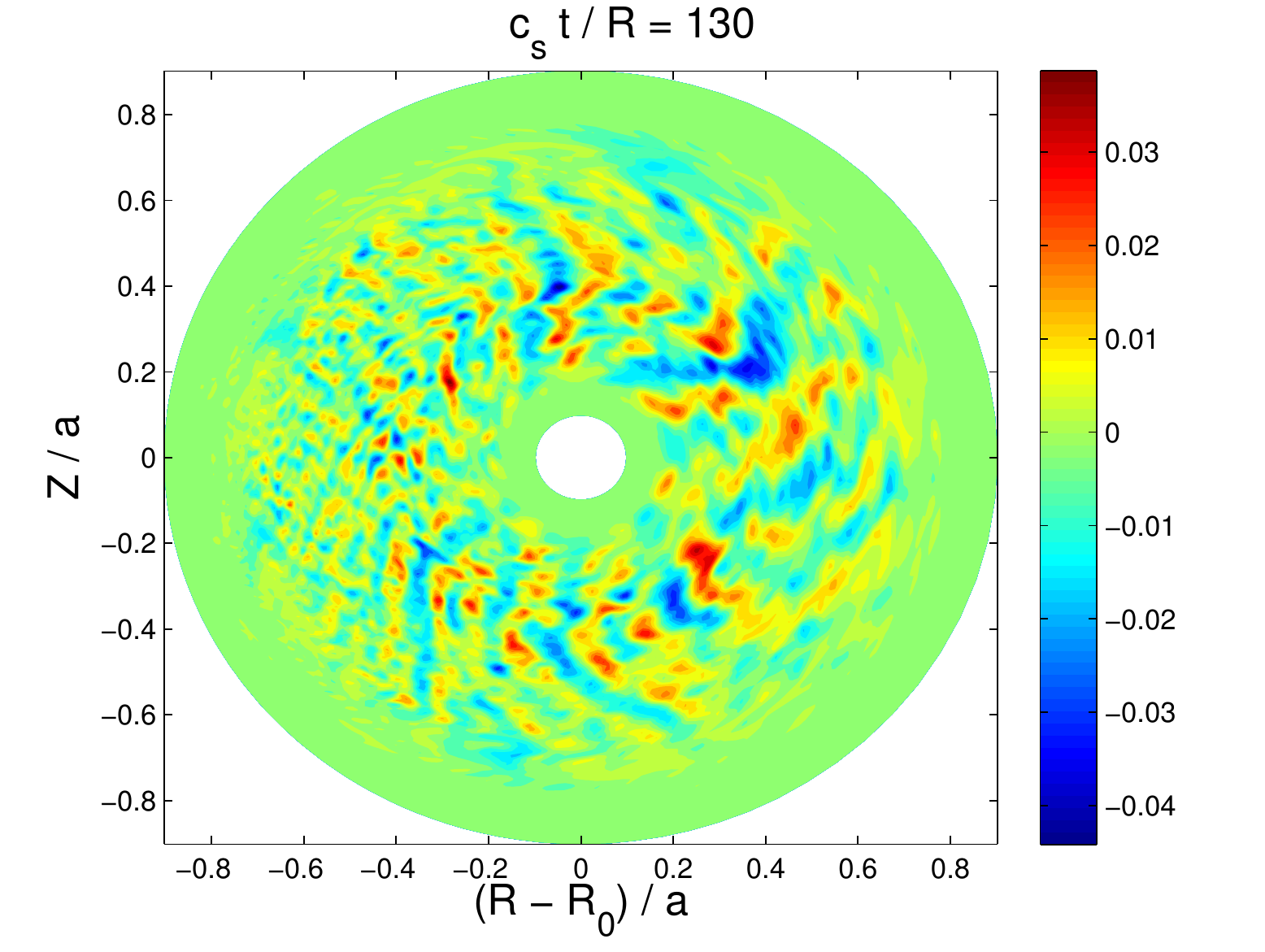}
\par\end{centering}
\caption{\label{fig:Structure}Contours of the non-zonal electrostatic potential
$\phi-\left\langle \phi\right\rangle _{FA}$ at $\frac{c_{s}t}{R}=130$.
Left and right figures are the results in NLT3 and NLT1, respectively.}
\end{figure}

Time evolutions of the ion heat diffusivity are shown in Fig.~(\ref{fig:kai}).
 It can be seen that the relaxation process obtained by using NLT
with magnetic axis is step-like, which is not observed in previous
simulation of relaxation process. Perturbation of the ion gyrocenter
center number with time is shown in Fig.~(\ref{fig:Number}). 
 The gyrocenter conservation is much improved by including the magnetic
axis in the simulation domain. The zonal field is shown in Fig.~(\ref{fig:phiZ}).
 Zonal fields obtained in NLT2 and NLT3 are similar with each other,
however, they are obviously different from the result obtained in
NLT1. The zonal electrostatic potential is nonzero, and the gradient
of the zonal electrostatic potential at the magnetic axis is almost
zero, which is reasonable. Fig.~(\ref{fig:Structure}) shows the
contours of the non-zonal electrostatic potential. 
The radial distribution of the perturbed electrostatic potential obtained
in simulation with magnetic axis is also different from that in simulation
without magnetic axis.

\section{Summary and discussion}

\label{Sec:8}

Simulation domain of the electrostatic gyrokinetic nonlinear turbulence
global code, NLT, is extended to include the magnetic axis in field-aligned
coordinates. In the first part of NLT for computing the unperturbed
guiding center orbit, Hamilton's equations in cylindrical coordinates
are solved when the guiding center is close to the magnetic axis,
thus the singularity of Poisson matrix in field-aligned coordinates
are avoided. Note that the unperturbed orbit is unchanged in NLT simulation,
which can be computed only once by using the high-precisional numerical
algorithm in the initial. The second part of NLT is the pull-back
transform, which is equivalent to compute the perturbed orbit. As
the method used in the first part, the pull-back transform at the
magnetic axis is computed by using formulations in cylindrical coordinates
for avoiding the singularity of Poisson matrix in field-aligned coordinates.
Numerically, partial derivatives in $R$ and $Z$ directions of cylindrical
coordinates are computed by using values at grid points $\left(\psi=\Delta_{\psi},\theta=-\pi\right)$,
$\left(\psi=\Delta_{\psi}\theta=0\right)$ and $\left(\psi=\Delta_{\psi},\theta=-\pi/2\right)$,
$\left(\psi=\Delta_{\psi},\theta=\pi/2\right)$ in field-aligned coordinates
and combining the toroidal Fourier transform. All these four points
are grid points in the $\psi-\theta$ plane. Thus, the coordinate
transform from field-aligned coordinates to cylindrical coordinates
is not needed in NLT. The third part, Birdsell's method\citep{2004-Birdsell-Book}
is extended from 2D polar coordinates to 3D field-aligned coordinates
for solving the gyrokinetic quasi-neutrality equation in the long-wavelength
approximation with adiabatic electrons. The integral format is used
to discretized the equation, the boundary condition at the magnetic
axis is the natural boundary condition instead of a artificial boundary
condition. The zonal field is solved directly from the equation without
using the magenetic surface averaged equation. The coefficient matrix
of the discretized algebraic equation is a block tridiagonal matrix,
the tridiagonal matrix algorithm is used to reduce the computational
cost. In the fourth part, numerical filtering, a new condition for
limiting the shortwave in the $\theta^{*}$ direction,$\frac{m}{r}\rho_{s}<\left[k_{\theta}\rho\right]_{max}$,
is considered. At the region near the magnetic axis, the retained
toroidal mode number is reduced with the radius by considering this
new condition.

In the nonlinear ITG test, the gyrocenter conservation is much improved
by including the magnetic axis in the simulation domain. The zonal
field and the radial distribution of the perturbed electrostatic potential
are different from previous results without the magnetic axis.

The numerical algorithm for treating the self-consistent simulation
including the magnetic axis in field-aligned coordinates is presented
in this paper. Although the algorithm is used in numerical Lie transform
code, the key idea for treating the singularity and boundary condition
at magnetic axis can also be used in different simulation codes. In
the electromagnetic simulation, equation of Ampere's law is also in
the form same with Poisson's equation, should be numerically discretized
with integral format. 

\section*{Acknowledgements}

This work was supported by the National Natural Science Foundation
of China under Grant Nos. 11675176, 11775265, 11505240 and 11575246,
the National ITER program of China under Contract No. 2014GB113000
and the Fundamental Research Funds for the Central Universities under
Grant No. WK2030040092.

\appendix

\section*{Reference}


\end{document}